\documentclass[trackchanges,twocolumn,twocolappendix]{aastex701} 

\usepackage[version=4]{mhchem}
\usepackage{graphicx}
\usepackage{subcaption}
\usepackage{soul}
\usepackage{tabularx}
\usepackage{multirow}%
\usepackage{booktabs} 
\usepackage{seqsplit}
\usepackage{threeparttable}
\usepackage{bm}
\usepackage{placeins}   
\usepackage{float}      
\allowdisplaybreaks     
\usepackage{enumitem}

\begin{document}

\title{Photochemical Production of \ce{CS2} in Temperate-to-Warm Gas Giant Exoplanet Atmospheres}

\author[0000-0002-1551-2610]{Jeehyun Yang}
\affiliation{Department of Astronomy and Astrophysics, University of Chicago, Chicago, IL 60637, USA}
\email[show]{jeehyuny@uchicago.edu}  
\author[0000-0001-5909-4433]{Vighnesh Nagpal}
\altaffiliation{NSF Graduate Research Fellow}
\affiliation{Department of Astronomy and Astrophysics, University of Chicago, Chicago, IL 60637, USA}
\email{vnagpal@uchicago.edu}

\author[0000-0002-0659-1783]{Michael Zhang}
\altaffiliation{51 Pegasi b Fellow}
\affiliation{Department of Astronomy and Astrophysics, University of Chicago, Chicago, IL 60637, USA}
\email{}

\author[0000-0002-6215-5425]{Qiao Xue}
\affiliation{Department of Astronomy and Astrophysics, University of Chicago, Chicago, IL 60637, USA}
\email{}

\author[0000-0002-1337-9051]{Eliza M.-R. Kempton}
\affiliation{Department of Astronomy and Astrophysics, University of Chicago, Chicago, IL 60637, USA}
\email{ekempton@uchicago.edu}

\author[0000-0003-4733-6532]{Jacob L.\ Bean}
\affiliation{Department of Astronomy and Astrophysics, University of Chicago, Chicago, IL 60637, USA}
\email{jacobbean@uchicago.edu}

\author[0000-0001-6247-8323]{Michael R. Line}
\affiliation{School of Earth and Space Exploration, Arizona State University, Tempe, AZ 85287, USA}
\email{}

\author[orcid=0000-0002-9843-4354]{Jonathan J. Fortney}
\affiliation{Department of Astronomy and Astrophysics, University of California, Santa Cruz, 95064}
\email[]{jfortney@ucsc.edu}

\author[0000-0002-8518-9601]{Peter Gao}
\affiliation{Earth and Planets Laboratory, Carnegie Institution for Science, 5241 Broad Branch Road, NW, Washington, DC 20015, USA}
\email{pgao@carnegiescience.edu}

\author[0000-0001-8236-5553]{Matthew C.\ Nixon}
\altaffiliation{51 Pegasi b Fellow}
\affiliation{School of Earth and Space Exploration, Arizona State University, Tempe, AZ 85287, USA}
\email{mcnixon1@asu.edu}

\author[0000-0002-2875-917X]{Caroline Piaulet-Ghorayeb}
\altaffiliation{E. Margaret Burbidge Postdoctoral Fellow}
\affiliation{Department of Astronomy and Astrophysics, University of Chicago, Chicago, IL 60637, USA}
\email{}

\author[0000-0002-7352-7941]{Kevin B. Stevenson}
\affiliation{Johns Hopkins University Applied Physics Laboratory, 11100 Johns Hopkins Rd, Laurel, MD 20723, USA}
\email{kevin.stevenson@jhuapl.edu}

\author[0000-0003-2404-2427]{Madison Brady}
\affiliation{Department of Physics and Astronomy, Michigan State University, East Lansing, MI 48824, USA}
\email{bradym27@msu.edu}

\author[0000-0003-3191-2486]{Joost P. Wardenier}
\affiliation{Weltraumforschung und Planetologie, Physikalisches Institut, University of Bern, Gesellschaftsstrasse 6, 3012 Bern, Switzerland}
\email{joost.wardenier@unibe.ch}

\author[0000-0003-0156-4564]{Luis Welbanks}
\affiliation{School of Earth and Space Exploration, Arizona State University, Tempe, AZ 85287, USA}
\email{luis.welbanks@asu.edu}

\author[0000-0002-0875-8401]{Jean-Michel D\'esert}
\affiliation{Leibniz Institute for Astrophysics, AIP Potsdam, Potsdam, 14482 Potsdam, Germany}
\affiliation{DESY, Platanenallee 6, Zeuthen, D-15738, Germany}
\email[]{jmdesert@aip.de}

\author[0000-0002-3263-2251]{Guangwei Fu}
\affiliation{Department of Physics and Astronomy, Johns Hopkins University, Baltimore, MD, USA}
\email{guangweifu@gmail.com}

\author[0000-0001-9521-6258]{Vivien Parmentier}
\affiliation{Laboratoire Lagrange, Université de la Côte d’Azur, Observatoire de la Côte d’Azur, CNRS, Nice, France.}
\email[]{vivien.parmentier@oca.eu}

\author[0000-0002-4250-0957]{Diana Powell}
\affiliation{Department of Astronomy and Astrophysics, University of Chicago, Chicago, IL 60637, USA}
\email{diana.powell@uchicago.edu}



\begin{abstract}

Sulfur chemistry has emerged as an important probe of exoplanet atmospheres in the JWST era, although observational constraints have thus far been largely limited to \ce{SO2} and \ce{H2S} in warm and hot exoplanets. Recent JWST observations have revealed \ce{CS2} in several cooler gas-giant exoplanets, yielding a new tracer of sulfur chemistry. However, the detailed chemical pathways responsible for the formation of \ce{CS2} remain poorly understood. Here, we use TOI-6894~b, a temperate gas giant with evidence for \ce{CS2}, as a test case for one-dimensional photochemical kinetic-transport modeling and sensitivity analyses of \ce{CS2} chemistry. We show that \ce{CS2} is produced through coupled thermochemical and photochemical processes involving \ce{CH4} and \ce{H2S} as the primary carbon and sulfur reservoirs, with \ce{S2} photolysis driving disequilibrium sulfur chemistry. Our models provide a physically consistent explanation for the observed \ce{CS2} feature in TOI-6894~b. Extending our analysis to gas giant exoplanets spanning a wide range of $T_{\rm eq}$, we find that \ce{CS2} abundance peaks in temperate to warm atmospheres ($T_{\rm eq}\sim500$ -- 700\,K), and declines toward both lower and higher temperatures. This temperature dependence provides a unified framework for interpreting current \ce{CS2} observations, accounting for reported detections in temperate to warm planets and the lack of detections in colder and hotter giant exoplanets. Our results establish \ce{CS2} as a complementary probe of sulfur inventories and atmospheric metallicity in cool gas giant exoplanets.
\end{abstract}

\keywords{\uat{Astrochemistry}{75}, \uat{Exoplanet atmospheres}{487}, \uat{Extrasolar gas giants}{509}, \uat{Exoplanet atmospheric composition}{2021}}


\section{Introduction}
\label{sec:intro}

Sulfur-bearing molecules have long served as important probes of physical and chemical processes in a wide range of astrophysical environments and, in exoplanet atmospheres, provide valuable constraints on metallicity, vertical mixing, and photochemistry while also offering clues to planetary formation \citep{atreya2020deep, turrini2021tracing, pacetti2022chemical}. In the JWST era, detections of sulfur species such as \ce{SO2} \citep{Alderson-2023, Tsai_2023, Crossfield_2023, Powell_2024, welbanks_2024, dyrek2024so2, Crossfield_2025_SO2_shorline} and \ce{H2S} \citep{fu2024_h2s, Zhang_2025_platon,xuan2026compositions} have become increasingly common, establishing sulfur chemistry as an important component of atmospheric characterization.
More recently, evidence for carbon disulfide (\ce{CS2}) has been reported in several exoplanet atmospheres, including those of the warm Jupiter WASP-80~b \citep[$T_{\rm eq}\sim$830\,K;][]{triantafillides2026identification}, the inflated young sub-Neptune V1298~Tau~e \citep[$T_{\rm eq}\sim$530\,K;][]{dai2026photochemicalcs2gasdetected}, the temperate Jupiter TOI-6894~b \citep[$T_{\rm eq}\sim$420\,K;][]{zhang2026cnos}, and tentatively the sub-Neptune TOI-270~d \citep[$T_{\rm eq}$=387\,K;][]{Benneke-2024_jwst}. These discoveries suggest that sulfur-bearing carbon species may be more widespread in exoplanet atmospheres than previously recognized. 

The chemistry of sulfur-bearing molecules has also been extensively studied in the broader context of astrochemistry, including molecular clouds, protoplanetary disks, and the circumstellar envelopes of evolved stars, where molecular abundances provide important diagnostics of physical conditions, chemical evolution, and elemental inventories \citep[e.g.,][]{Agundez_2006,herbst2008chemistry,oberg2021astrochemistry,vanDishoeck2023}. In particular, sulfur-bearing species have long been recognized as valuable tracers of photochemical and nonequilibrium processes in these environments and have been used to investigate the long-standing sulfur depletion problem and the sulfur budget in planetary systems \citep[e.g.,][]{millar1990organo, vidal2017reservoir, laas2019modeling, Reed_2024}.

Unlike the more extensively studied sulfur species \ce{SO2} and \ce{H2S}, the chemical pathways responsible for the formation of \ce{CS2} and the atmospheric conditions that favor its production remain poorly understood.
Previous studies that considered \ce{CS2} formation pathways in cool giant exoplanets \citep{Zahnle_2009, mukherjee2025effects} have generally focused on a small number of key intermediates, such as \ce{H2CS} \citep{moses2024sulfur,Veillet-2026,triantafillides2026identification,dai2026photochemicalcs2gasdetected}. However, the relative importance of competing chemical pathways and the sensitivity of \ce{CS2} abundances to model parameters and chemical networks have not been systematically quantified, motivating further investigation of the underlying chemistry.

Among the gas-giants with \ce{CS2} detections, TOI-6894~b provides an especially valuable opportunity to investigate \ce{CS2} chemistry. TOI-6894~b is a temperate giant exoplanet ($R=0.86 \,R_J$, $M=0.17\,M_J$) orbiting an M5 dwarf with a period of 3.4 days \citep[$R_*=0.23\,R_\odot, M_*=0.21\,M_\odot, T_{\rm eff}=3007\,{\rm K}$;][]{bryant2025transiting}. The system presents compelling scientific opportunities. For example, TOI-6894~b belongs to an emerging population of giant planets orbiting very low-mass stars, whose existence presents significant challenges to conventional core accretion models of giant planet formation \citep{morales2019giant}. Furthermore, the exceptionally large $R_p/R_* = 0.39$ produces transmission spectral features of approximately 1,300 ppm per scale height and a large transmission spectroscopy metric \citep[TSM = 450 at 3 $\mu$m;][]{kempton2018framework}, making TOI-6894~b one of the most favorable targets for constraining the atmospheric inventories of carbon, nitrogen, oxygen, and sulfur from a single JWST observation.

TOI-6894~b was recently observed with JWST using the NIRSpec/PRISM mode (GO 8696, PI: Michael Zhang). The observations revealed a distinct absorption feature peaking at 4.6 $\mu$m that could not be reproduced by thermochemical equilibrium models. An extensive search for candidate absorbers identified \ce{CS2} as the most likely cause of the observed 4.6 $\mu$m feature \citep{zhang2026cnos}. The identification of \ce{CS2} in TOI-6894~b therefore provides an opportunity to investigate its origin and to its broader role in exoplanet sulfur chemistry.

In this Letter, we investigate the photochemical production of \ce{CS2} in giant exoplanet atmospheres. We perform forward modeling of temperate-to-warm giant exoplanet atmospheres, including calculations of atmospheric thermal structures and atmospheric chemical abundance profiles. Using TOI-6894~b as a case study as well as a broader sample of giant exoplanets spanning a wide range of $T_{\rm eq}$, we explore the conditions under which \ce{CS2} forms, identify its dominant chemical pathways, and demonstrate that \ce{CS2} represents an important sulfur-bearing species in disequilibrium atmospheric chemistry.

\section{Methods} \label{sec:methods}
We adopted a one-dimensional, globally averaged framework (using \texttt{PICASO} for the $T$-$P$ structure and \texttt{EPACRIS} for the photochemical calculations) because the primary goal of this study is to investigate the chemical pathways governing \ce{CS2} formation and their dependence on key atmospheric parameters, rather than to reproduce the full three-dimensional atmospheric structure. Such one-dimensional models have been widely used to interpret transmission spectra and to study atmospheric chemistry in exoplanets, as transmission spectroscopy primarily probes the terminator-averaged atmospheric structure.

\subsection{Atmospheric $T$-$P$ Profile Modeling (\texttt{PICASO})} \label{subsec:picaso}

We used \texttt{PICASO 4.0} \citep{mang2026picaso} to perform one-dimensional radiative--convective equilibrium (RCE) calculations and generate self-consistent, planet-wide average $T$--$P$ profiles for TOI-6894~b assuming full day--night heat redistribution\footnote{This corresponds to \texttt{rfacv = 0.5} \citep{mukherjee_picaso2023}}. The models span 91 pressure levels ranging from $10^{-6}$ to $10^{3}$~bar. We computed model grids spanning five atmospheric metallicities \citep[0.1, 1, 3, 10, and 30 $\times Z_{\odot}$, where $Z_{\odot}$ denotes the solar elemental abundance ratios with C/O = 0.55 adopted from][]{lodders2021relative} and five internal temperatures ($T_{\mathrm{int}}$ = 30, 60, 100, 200, and 400~K). The adopted metallicity range spans subsolar to substantially supersolar compositions and encompasses the values commonly inferred for giant exoplanets and predicted by planet formation models \citep{fortney2013framework, thorngren2016mass}, while also allowing us to investigate the sensitivity of \ce{CS2} chemistry to elemental enrichment. The adopted $T_{\mathrm{int}}$ values bracket the range expected for both mature giant planets and younger, more intrinsically luminous systems, allowing us to assess the sensitivity of \ce{CS2} production to deep atmospheric heating.

The incident flux upon TOI-6894~b was calculated through interpolating the \texttt{PHOENIX} \citep{allard-2012-phoenix} grid of models according to the host star's stellar parameters, as described in \citet{mukherjee_picaso2023}. We note that the RCE and photochemical calculations (described in Section~\ref{subsec:epacris}) were not self-consistently coupled. Previous work shows that disequilibrium chemistry has only a modest effect on the thermal structure of gas giant atmospheres with metallicities below $10\times Z_{\odot}$ \citep{mukherjee2025effects}; therefore, we do not expect this approximation to significantly affect our results.

\subsection{Chemical Networks} \label{subsec:chemical_network}

The photochemical network adopted in this study consists of two components. The primary chemical network was adopted from \citet{yang2024chemical} (\url{https://doi.org/10.5281/zenodo.16750016}), which was generated using the automated Reaction Mechanism Generator \citep[\texttt{RMG};][]{Gao_2016, rmg-v3, RMG-database}, a mature and extensively benchmarked framework originally developed for combustion chemistry.  Briefly, \texttt{RMG} follows an automated algorithm for identifying which chemical species and reactions are relevant under specified physical conditions. \texttt{RMG}'s extensive validation against experimental measurements and high-level theoretical calculations provides confidence in its generation of chemically comprehensive reaction networks. The \texttt{RMG} framework has been applied successfully both in the exoplanet context \citep{yang2024epacris, yang2024chemical, Bello-Arufe_2026, baeyens2026phasedependentchemistrywasp43b} and to solar system planets \citep{yang2026jupiter} --- notably Jupiter, which \citet{zhang2026cnos} found to have similar abundance patterns to TOI-6894~b.
 
Unfortunately, the \texttt{RMG} reaction network does not include \ce{CS2}, which initially limited our ability to model the observed properties of TOI-6894~b. We therefore augmented the network with sulfur-bearing species and their associated reactions that were absent from the network but available in the \texttt{VULCAN} chemical network \citep{Tsai_2017, Tsai_2021}, specifically \texttt{SNCHO\_photo\_network\_2025.txt} \citep{VULCAN_SNCHO_2025}. This included the species \ce{CS}, \ce{CS2}, \ce{H2CS}, \ce{HCS}, and \ce{NS}, together with all associated photochemical and thermochemical reactions available in that network (3 photochemical and 36 thermochemical reactions). Thus, only a small fraction of the full \texttt{SNCHO\_photo\_network\_2025} network was incorporated, while the underlying reaction mechanism remained the \texttt{RMG}-generated network. The resulting chemical network contains 103 species, including 33 sulfur-bearing species, and 2067 reactions. Further details of the chemical network are provided in Appendix~\ref{sec:appendix_chemical_network}.

\subsection{One-dimensional Photochemical Kinetic-Transport Atmospheric Modeling (\texttt{EPACRIS})} \label{subsec:epacris}

We modeled the atmosphere of TOI-6894~b using the one-dimensional photochemical kinetic-transport code \texttt{EPACRIS} \citep{yang2024epacris}. For the stellar irradiation input, we adopted the spectrum of the M5-type star GJ~876 from the \texttt{MUSCLES} survey III \citep[$T_{\rm eff}=3062$\,K;][]{loyd_2016}, whose effective temperature is similar to that of TOI-6894 \citep[$T_{\rm eff}=3007$\,K;][]{bryant2025transiting}. The spectrum was scaled to reproduce the bolometric insolation received by TOI-6894~b (5.54 $S_{\oplus}$; \citealt{bryant2025transiting}). The photochemical modeling depends sensitively on the stellar ultraviolet flux, whereas the RCE calculations are primarily controlled by the bolometric irradiation. We therefore adopted the empirically constrained \texttt{MUSCLES} spectrum of GJ~876, which provides a more realistic UV spectrum than the corresponding \texttt{PHOENIX} models for M dwarfs.

According to the companion observational study \citep{zhang2026cnos}, the inferred atmospheric metallicity of TOI-6894~b lies between 3$\times$ and 10$\times Z_{\odot}$. We therefore adopt a fiducial atmospheric model with a metallicity of 3$\times Z_{\odot}$ (We adopt the solar elemental abundances, $Z_{\odot}$, from Table~3 of \citet{lodders2021relative}. The atmospheric metallicities considered in this work are defined as uniform scalings of these solar elemental abundances while preserving the solar elemental ratios), a uniform eddy diffusion coefficient of $K_{\rm zz}=10^8$ cm$^2$ s$^{-1}$, and $T_{\rm int}=100$\,K. The latter two parameters are representative values of Jupiter's deep atmosphere commonly adopted for gas-giant atmospheres \citep{BJORAKER1986579, li2012emitted}. Because vertical mixing in exoplanet atmospheres remains poorly constrained, we adopt a vertically uniform eddy diffusion coefficient, a common approximation in one-dimensional photochemical studies. Our primary goal is to investigate the sensitivity of \ce{CS2} chemistry to the overall efficiency of vertical transport.

The fiducial model described above serves as the baseline case for TOI-6894~b. In the following section, we explore variations in metallicity, $K_{zz}$, $T_{\rm int}$, UV irradiation, and the chemical network to assess the sensitivity of \ce{CS2} production to these parameters and to test the robustness of our interpretation of the observations \citep{zhang2026cnos}.

\subsection{Sensitivity Analysis} \label{subsec:method_sensitivity_analysis}
To assess the sensitivity of the atmospheric chemistry of TOI-6894~b, we computed an additional grid of models using \texttt{PICASO} (Section \ref{subsec:picaso}) and \texttt{EPACRIS} (Section \ref{subsec:epacris}), each with its own radiative-convective equilibrium $T$--$P$ profile, exploring $K_{zz}$ ranging from $10^6$ to $10^{10}$ cm$^2$ s$^{-1}$, metallicities from 0.1$\times$ to 30$\times Z_{\odot}$, and $T_{\rm int}$ spanning 30 to 400\,K. We also adopted the stellar spectrum of AD Leo, an active M4.5 dwarf from the \texttt{MUSCLES} survey III \citep[$T_{\rm eff}=3400$ K;][]{loyd_2016}, scaling it to the same bolometric insolation as TOI-6894~b to assess the impact of enhanced UV irradiation. In these sensitivity tests, $K_{\rm zz}=10^8$ cm$^2$ s$^{-1}$, [M/H] = 0 (solar metallicity), $T_{\rm int}=100$\,K, and the GJ~876 stellar spectrum were adopted as the reference model.

Chemical kinetic models inherently contain uncertainties arising from imperfect knowledge of reaction rate coefficients. These uncertainties propagate into the model output and therefore represent one of the major caveats in atmospheric chemical modeling. Owing to decades of developments in combustion chemistry, our understanding of the chemistry of C-H-O species, and to some extent nitrogen chemistry, has been significantly improved through laboratory measurements of key reactions and \textit{ab initio} calculations \citep[e.g.,][]{green2007predictive, kohse2021combustion, miller2021combustion}. However, sulfur chemistry remains comparatively poorly constrained because of its chemical complexity and experimental difficulties \citep{STAGNI2022136723}. For example, sulfur-bearing species are often highly reactive and can contaminate or damage chemical kinetic measurement systems (e.g., reactor walls and mass spectrometers), making laboratory studies particularly challenging. For this reason, sensitivity analysis is routinely performed in chemical engineering applications to evaluate the robustness of chemical kinetic models against uncertainties in reaction rate coefficients. Such analyses not only assess the reliability of model predictions but also identify the reactions that exert the strongest control over the chemical system.

Accordingly, we evaluated the sensitivity of the model to rate coefficients associated with the major \ce{CS2} formation pathways using the fiducial model (3$\times Z_{\odot}$, $K_{zz}=10^8$ cm$^2$ s$^{-1}$). We first identified the reactions that significantly contribute to \ce{CS2} production and loss at $P\sim0.1$ mbar in the atmosphere of TOI-6894~b, corresponding to the pressure region of primary interest for both photochemistry and JWST observations. We then perturbed each reaction individually by reducing its rate coefficient (i.e., $k$) by a factor of 10 and performed 1D photochemical modeling to examine the resulting response in the atmospheric \ce{CS2} abundance at $P\sim0.1$ mbar (i.e., [\ce{CS2}]). This approach quantifies how uncertainties in the model output (here, [\ce{CS2}]) propagate from uncertainties in the model input (here, $k$). The sensitivity is expressed using the normalized \ce{CS2} concentration sensitivity with respect to the rate coefficients, hereafter referred to as the sensitivity coefficient:
\begin{equation}
\label{eqn:sensitivity_coefficient}
    S_{i,\ce{CS2}} = \frac{d(\ln[\ce{CS2}])}{d(\ln k_i)}
\end{equation}
\noindent where $i$ denotes reaction $i$. We calculated the sensitivity coefficients of the \ce{CS2} abundance with respect to 27 reaction rate coefficients identified as the major contributors to \ce{CS2} formation.

\subsection{Transmission Spectra Modeling (\texttt{PLATON})} \label{subsec:platon}
After the 1D photochemical models reached steady state, we computed synthetic transmission spectra of TOI-6894~b using the resulting vertical molecular mixing ratio profiles with \texttt{PLATON} \citep{Zhang_2019_platon}. In this work, we used \texttt{PLATON} v6.2 \citep{Zhang_2025_platon}.  We added \ce{CS2} opacities to the default opacity library from \cite{Zhang_2019_platon} using line lists for all isotopologues from the \texttt{HITRAN} database \citep{GORDON_2026_Hitran}. The line lists span wavenumbers from 64 to 7550 cm$^{-1}$ (1.3--156.3 $\mu$m). From these line lists \citep{GORDON_2026_Hitran}, we computed wavelength-dependent absorption cross sections on the standard \texttt{PLATON} opacity grid, covering temperatures from 100 to 4000 K in 100 K increments, pressures from $10^{-4}$ to $10^{8}$ Pa in decade increments, and wavelengths from 0.2 to 30 $\mu$m at a resolving power of $R=20,000$. The modeled spectra were then compared to the JWST observations \citep{zhang2026cnos}. Because our calculations employ one-dimensional atmospheric profiles, the simulated transmission spectra correspond to disk-integrated spectra of a horizontally homogeneous atmosphere and do not account for day-night or limb inhomogeneities.

\subsection{Exploring the Dependence of \ce{CS2} Formation on equilibrium temperature ($T_{\rm eq}$)} \label{subsec:teq_dependence_method}
Together with recent \ce{CS2} detections in V1298~Tau~e \citep[$T_{\rm eq}\sim530$\,K;][]{dai2026photochemicalcs2gasdetected} and WASP-80~b \citep[$T_{\rm eq}\sim830$\,K;][]{triantafillides2026identification}, the detection of \ce{CS2} in TOI-6894~b ($T_{\rm eq}\sim420$\,K) raises the question of how \ce{CS2} production depends on $T_{\rm eq}$. Motivated by the ability of our photochemical framework to reproduce the atmospheric properties of TOI-6894~b, we extended our analysis to five additional temperate-to-warm gas giant exoplanets that have already been well-characterized with JWST: TOI-199~b \citep{Bello-Arufe_2026}, V1298~Tau~e \citep{dai2026photochemicalcs2gasdetected}, V1298~Tau~b \citep{barat2025metal}, WASP-80~b \citep{triantafillides2026identification}, and WASP-39~b \citep{Rustamkulov-2023}. Together, these planets span a broad range of equilibrium temperatures ($T_{\rm eq}=350$--1120\,K) and include both detections and non-detections of \ce{CS2}. For each planet, we performed \texttt{EPACRIS} simulations following the methodology described in Section~\ref{subsec:epacris} and considered both 1$\times$ and 10$\times Z_{\odot}$ metallicities. Details of these models are provided in Appendix~\ref{sec:other_exoplanets}.

\begin{figure*}[ht!]
\centering
\includegraphics[width=0.75\textwidth]{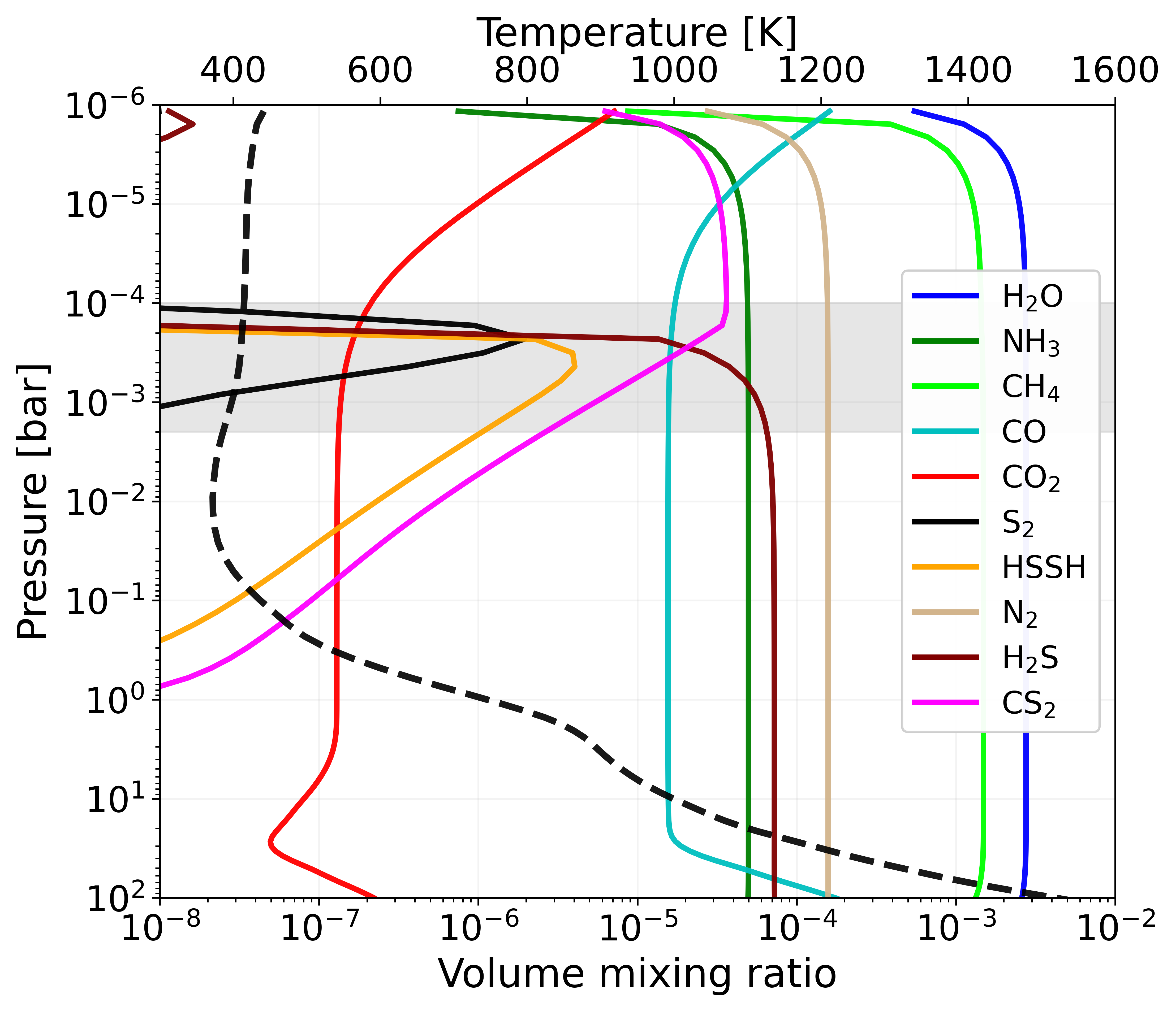}
\caption{Vertical volume mixing ratio profiles for the fiducial atmospheric model of TOI-6894~b ($3\times Z_{\odot}$ metallicity, $K_{\rm zz}=10^8$ cm$^2$ s$^{-1}$, and $T_{\rm int}=100$\,K). Solid colored lines show the volume mixing ratio profiles of individual chemical species, while the dashed black line represents the corresponding $T$--$P$ profile computed using \texttt{PICASO} as described in Section~\ref{subsec:picaso}. \ce{SO2} is not shown because its VMR lies below the plotted range. The grey shaded region indicates the approximate pressure range probed by JWST transmission spectroscopy \citep[0.1--2 mbar;][]{Rustamkulov-2023}.}
\label{fig:3x_vmr}
\end{figure*}

\begin{figure*}[htb!]
    \includegraphics[width=1\textwidth]{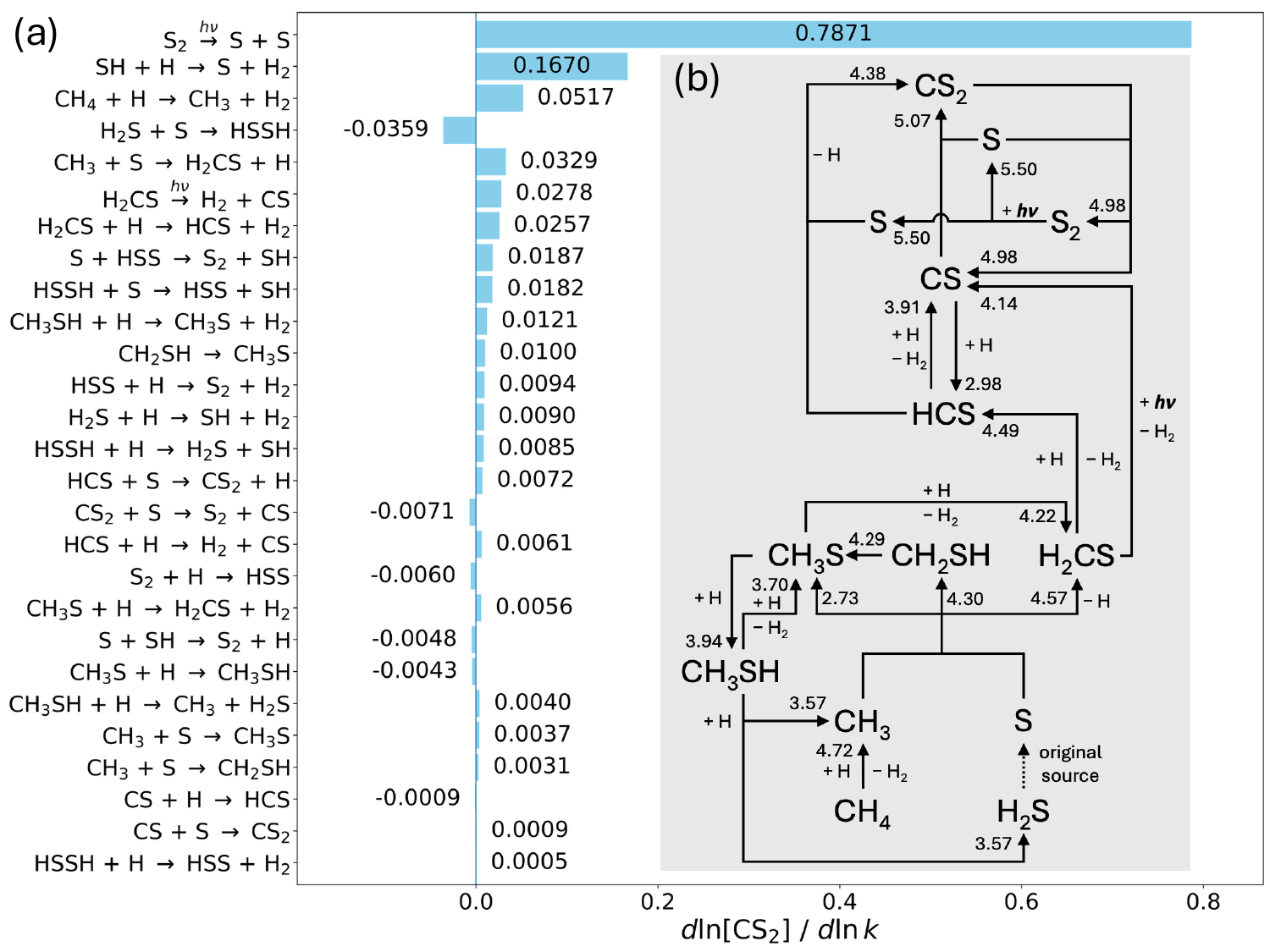}
     \caption{\footnotesize (a) Sensitivity coefficients ($S_{i,\mathrm{\ce{CS2}}}=\frac{d(\rm ln[\mathrm{CS_2}])}{d(\rm ln \textit{k}_i)}$) for the major reactions contributing to \ce{CS2} formation at $P\sim0.1$ mbar, shown in descending orders from top to bottom. Briefly speaking, reactions with large positive (negative) sensitivity coefficients promote (suppress) \ce{CS2} formation (see Section~\ref{subsec:method_sensitivity_analysis} for the definition and Section~\ref{sec:sensitivity_analysis_rate_coefficients} for further discussion). Among these, \ce{S2} photolysis is identified as the most influential reaction. (b) A schematic diagram illustrating major reaction pathways contributing to the formation of \ce{CS2} at the $P\sim0.1$ mbar region in the simulated atmosphere of TOI-6894~b, under the fiducial model conditions ($3\times Z_{\odot}$, $K_{\rm zz}=10^8$ cm$^2$ s$^{-1}$, and $T_{\rm int}=100$\,K), corresponding to the results shown in Figure~\ref{fig:3x_vmr}. Numbers indicate the logarithm of the absolute reaction rate, in units of molecules cm$^{-3}$ s$^{-1}$, for each reaction pathway. For example, a value of 5.50 for the \ce{S2} destruction pathway corresponds to a rate of 10$^{5.50}$ molecules cm$^{-3}$ s$^{-1}$ for the reaction \ce{S2}$\xrightarrow{h\nu}$S+S reaction. Note that some reactions whose sensitivity coefficients are calculated in panel (a) are omitted from the schematic diagram in panel (b) for visual clarity. As shown, \ce{S2} photolysis, which efficiently liberates two sulfur atoms, plays a key role in \ce{CS2} formation.}
    \label{fig:pathways_sensitivity}
\end{figure*}

\section{Results and Discussions} \label{sec:results_discussions}
\subsection{Overall Atmospheric Chemistry of TOI-6894~b} \label{sec:overall_chemistry}

Figure~\ref{fig:3x_vmr} shows the vertical volume mixing ratio (VMR) profiles for the fiducial TOI-6894~b model described in Section~\ref{subsec:epacris}. In the deep atmosphere ($P\geq10$ bar), the C-, N-, O-, and S-bearing species are present in their typical forms expected for a \ce{H2}-dominated atmosphere and are readily quenched. Oxygen is primarily locked in \ce{H2O}, carbon in \ce{CH4}, nitrogen in both \ce{N2} and \ce{NH3}, and sulfur in \ce{H2S}. This behavior reflects the dominance of thermochemical equilibrium in the deep atmosphere, where high temperatures lead to thermochemical timescales much shorter than vertical mixing or photochemical timescales \citep{moses2014chemical}. In the middle atmosphere ($P\sim10^{-2}-1$ bar), vertical mixing controls the overall composition, producing nearly uniform VMR profiles for the major C-, N-, O-, and S-bearing species. The nearly constant abundance profiles of species such as \ce{CH4}, \ce{CO}, \ce{CO2}, \ce{N2}, \ce{NH3}, \ce{H2O}, and \ce{H2S} are characteristic of transport-induced quenching and indicate that vertical mixing dominates over local chemistry. In the upper atmosphere ($P\lesssim10^{-2}$ bar), photochemistry controls the overall composition. For example, starting from the gray-shaded region in Figure~\ref{fig:3x_vmr}, \ce{H2S} and \ce{CO2} depart substantially from their quenched abundances and exhibit strong altitude-dependent abundance gradients, indicative of active photochemical processing. Near the upper boundary, the decline in species such as \ce{H2O} and \ce{CH4} also reflects photochemical destruction.

A key result is the formation of \ce{CS2} as the dominant sulfur-bearing species in the observable upper atmosphere, represented by the magenta solid line in Figure~\ref{fig:3x_vmr}. Within the grey-shaded JWST-observable region \citep[$P\sim0.1$--2 mbar;][]{Rustamkulov-2023}, sulfur initially stored as \ce{H2S} in the deep atmosphere is progressively converted primarily into \ce{CS2} toward higher altitudes, with smaller fractions forming \ce{S2} and \ce{HSSH}. Consequently, the upper-atmosphere VMR of \ce{CS2} is approximately a factor of two lower than the deep-atmosphere VMR of \ce{H2S}, since each \ce{CS2} molecule contains two sulfur atoms, whereas each \ce{H2S} molecule contains only one. This result indicates that sulfur originally locked in \ce{H2S} in the deep atmosphere is largely redistributed into \ce{CS2} in the upper atmosphere of TOI-6894~b. In the next section, we further identify the production pathways of \ce{CS2}.

\subsection{Chemical Formation Pathways of \ce{CS2}} \label{sec:cs2_formation_pathways}

Figure~\ref{fig:pathways_sensitivity}a shows the sensitivity coefficients of the major reactions influencing \ce{CS2} formation at $P\sim0.1$ mbar in Figure~\ref{fig:3x_vmr}, while Figure~\ref{fig:pathways_sensitivity}b presents the dominant reaction pathways from \ce{H2S} to \ce{CS2} in the same region. Since \ce{CS2} contains both carbon and sulfur atoms, its formation requires carbon and sulfur reservoirs. In the atmosphere of TOI-6894~b, the dominant carbon source is \ce{CH4}, while the dominant sulfur source is \ce{H2S}.

According to the steady-state atmospheric solution, \ce{CH4} reacts with H atoms to produce \ce{CH3} radicals and \ce{H2}. Sulfur atoms are liberated from \ce{H2S} and subsequently maintained through a reaction network involving \ce{H2}, H, \ce{HSS}, \ce{HSSH}, SH, and \ce{S2}. The resulting \ce{CH3} and S radicals react to produce three channels: \ce{CH3S}, \ce{CH2SH}, and \ce{H2CS}+H (see Figure~\ref{fig:pathways_sensitivity}b). Among these, the \ce{CH2SH} and \ce{H2CS}+H channels dominate, with branching ratios of 34.6\% and 64.5\%, respectively, whereas the \ce{CH3S} channel contributes only 0.9\%. Most \ce{CH2SH} subsequently isomerizes into \ce{CH3S}. The \ce{CH3S} radical then reacts with H atoms to produce either \ce{H2CS} (65.6 \%) or \ce{CH3SH} (34.4 \%). The formation of \ce{H2CS} is particularly important because it acts as the key intermediate toward \ce{CS2} production. The importance of \ce{H2CS} as an intermediate for \ce{CS2} formation has also been pointed out by previous conference work exploring sulfur chemistry in warm sub-Neptunes \citep{moses2024sulfur}, as well as in recent modeling and observational studies \citep{Veillet-2026,triantafillides2026identification,dai2026photochemicalcs2gasdetected}. About 69.1\% of \ce{H2CS} reacts with H atoms to form HCS, which subsequently reacts with H atoms again and produces CS+\ce{H2}. The remaining 30.9\% \ce{H2CS} undergoes photolysis directly to form CS+\ce{H2}. Sulfur atoms then react with either CS or HCS to form \ce{CS2}, although some go through reactions with \ce{CS2} to form \ce{S2}. Overall, this sequence represents the dominant pathway for \ce{CS2} formation in TOI-6894~b.

As briefly mentioned in Section~\ref{subsec:chemical_network}, the original \texttt{Photochem} network failed to produce sufficient \ce{CS2} within the JWST-observable region, predicting \ce{CS2} mixing ratios of only $\sim10^{-7}$
 near the 1 mbar level and $\sim10^{-6}$ at most at 0.1 mbar. Unlike the other chemical networks considered in this study, the \texttt{Photochem} \citep{wogan2025photochem} network does not include \ce{H2CS}, \ce{CH3S}, or \ce{CH3SH}, because it was primarily developed for early Earth atmospheric applications, which are generally assumed to be \ce{CO2}- or \ce{N2}-dominated \citep{holland2020chemical, catling2017atmospheric}. Under such oxidizing conditions, reduced sulfur species such as \ce{CH3SH}, \ce{CH3S}, and \ce{H2CS} are chemically less important and were therefore omitted from the network. However, under \ce{H2}-dominated and low-metallicity gas-giant conditions, species such as \ce{H2CS}, \ce{CH3S}, and \ce{CH3SH} can become important sulfur reservoirs and intermediates linking carbon and sulfur chemistry. The omission of \ce{H2CS} may also explain why neither the chemical networks of \citet{wogan2025photochem} nor that of \citet{yang2024chemical} predicted significant \ce{CS2} abundances in the atmosphere of TOI-270~d \citep{Benneke-2024_jwst}. However, the current TOI-270~d observation \citep{Benneke-2024_jwst} remains too uncertain in the NIRSpec wavelength region beyond 4.5~$\mu$m to conclusively attribute the reported feature to \ce{CS2} rather than overlapping CO absorption, particularly given the highly oxidizing nature of the atmosphere inferred from its high metallicity \citep[$\geq230\times Z_{\odot}$;][]{Benneke-2024_jwst}. This uncertainty strongly motivates additional and more detailed observations of TOI-270~d.

\subsection{Sensitivity Analysis of \ce{CS2} Formation to Rate-Coefficient Uncertainty} \label{sec:sensitivity_analysis_rate_coefficients}

Based on the dominant reactions involved in \ce{CS2} formation identified through the rate analysis described in Figure~\ref{fig:pathways_sensitivity}b and Section~\ref{sec:cs2_formation_pathways}, we subsequently performed sensitivity calculations for each reaction following the procedure described in Section~\ref{subsec:method_sensitivity_analysis}. The resulting sensitivity coefficients are shown in Figure~\ref{fig:pathways_sensitivity}a. The model sensitivity is quantified using the sensitivity coefficient (Eqn.~\ref{eqn:sensitivity_coefficient}). Briefly speaking, if the sensitivity coefficient of the \ce{CS2} abundance with respect to reaction $i$ is equal to 1 (i.e., $S_{i,\ce{CS2}}$=1), reducing the rate coefficient of reaction $i$ by a factor of 10 results in a corresponding factor of 10 decrease in the predicted \ce{CS2} abundance, indicating that the reaction positively contributes to \ce{CS2} production. Conversely, if $S_{i,\ce{CS2}}=-1$, the reaction destroys \ce{CS2}: decreasing the rate coefficient by a factor of 10 increases the predicted \ce{CS2} abundance by a factor of 10, whereas increasing the rate coefficient decreases the \ce{CS2} abundance. In practice, reactions with small absolute sensitivity coefficients have a limited influence on the model output. For example, $|S_{i,\ce{CS2}}|\leq0.1$ implies that even a factor of 10 perturbation in the rate coefficient changes the predicted \ce{CS2} abundance by only $\sim$20 \%, while $|S_{i,\ce{CS2}}|\leq0.01$ corresponds to only $\sim$2 \% variation.

Figure~\ref{fig:pathways_sensitivity}a shows that among the 27 dominant reactions associated with \ce{CS2} formation, 25 reactions possess $|S_{i,\ce{CS2}}|\lesssim0.05$, corresponding to uncertainties of only $\lesssim25$ \% in the predicted \ce{CS2} abundance even under order-of-magnitude perturbations to their individual rate coefficients. The SH+H$\rightarrow$S+\ce{H2} reaction exhibits the second largest sensitivity ($S_{i,\ce{CS2}}=0.1670$), corresponding to approximately a factor of two variation in \ce{CS2} abundance even if the reaction rate is perturbed by two orders of magnitude. Overall, these results demonstrate the robustness of the predicted \ce{CS2} production against uncertainties in most reaction rates, noting that the order-of-magnitude perturbations adopted here are conservative for many kinetic parameters constrained theoretically or experimentally.

The only truly influential reaction is \ce{S2} photolysis,
\begin{equation}
    \ce{S2}\xrightarrow{h\nu}\ce{S}+\ce{S}
\end{equation}

\noindent whose sensitivity coefficient approaches unity ($S_{i,\ce{CS2}}=0.7871$). We verified that this conclusion remains unchanged throughout the JWST-observable region ($P\sim0.1$--2 mbar): \ce{S2} photolysis remains the only one dominant controlling process governing \ce{CS2} production, although the identity of the secondary reactions changes somewhat with pressure. This indicates that \ce{S2} photolysis acts as the primary controlling reaction for \ce{CS2} production in the JWST-observable atmosphere of TOI-6894~b, suggesting that \ce{CS2} formation is primarily photochemically driven. Consequently, the abundance of \ce{CS2} is particularly sensitive to the stellar flux between $\sim240$ and 280 nm \citep{green1996deperturbation, green1997upper, stark2018fourier}, which corresponds to the major absorption band of \ce{S2} photolysis. 

\ce{S2} photolysis proceeds through a predissociation mechanism. Briefly speaking, predissociation occurs when dissociation takes place at energies below the nominal dissociation threshold due to strong spin-orbit coupling, which induces state mixing between bound and dissociative electronic states. The ground state of \ce{S2} ($X^3\Sigma^-_g$) is first excited into its quantum mechanically allowed bound excited state $B^3\Sigma^-_u$. However, the strong spin-orbit coupling in \ce{S2} causes significant state mixing between $B^3\Sigma^-_u$ and its quantum mechanically forbidden state $B^{''3}\Pi^-_u$. Since $B^{''3}\Pi^-_u$ possesses a substantially lower dissociation threshold ($\sim280$ nm) than $B^3\Sigma^-_u$ ($\sim230$ nm), state mixing enables photodissociation of \ce{S2} at significantly longer wavelengths.

This $B/B^{''}$--$X$ system of \ce{S2} has been extensively investigated both experimentally \citep{green1996deperturbation, green1997upper, stark2018fourier} and theoretically \citep{hull2020collisional}. In \texttt{EPACRIS}, we adopt the \ce{S2} photodissociation cross sections from \citet{hrodmarsson2023photodissociation}, which were derived from experimental measurements by \citet{stark2018fourier} at 370\,K over the wavelength range 241.0--281.0 nm, with estimated uncertainties below 30\%. Therefore, even under conservative assumptions, the uncertainty in the \ce{S2} photolysis rate is unlikely to exceed a factor of two. Combining this with the sensitivity coefficient of the \ce{S2} photolysis reaction ($S_{i,\ce{CS2}}=0.7871$), the predicted \ce{CS2} abundance would vary by at most $\sim$40\%. This result further supports the robustness of the photochemically produced \ce{CS2} abundance in the modeled atmosphere of TOI-6894~b.

\begin{figure}[htb!]
    \includegraphics[width=0.43\textwidth]{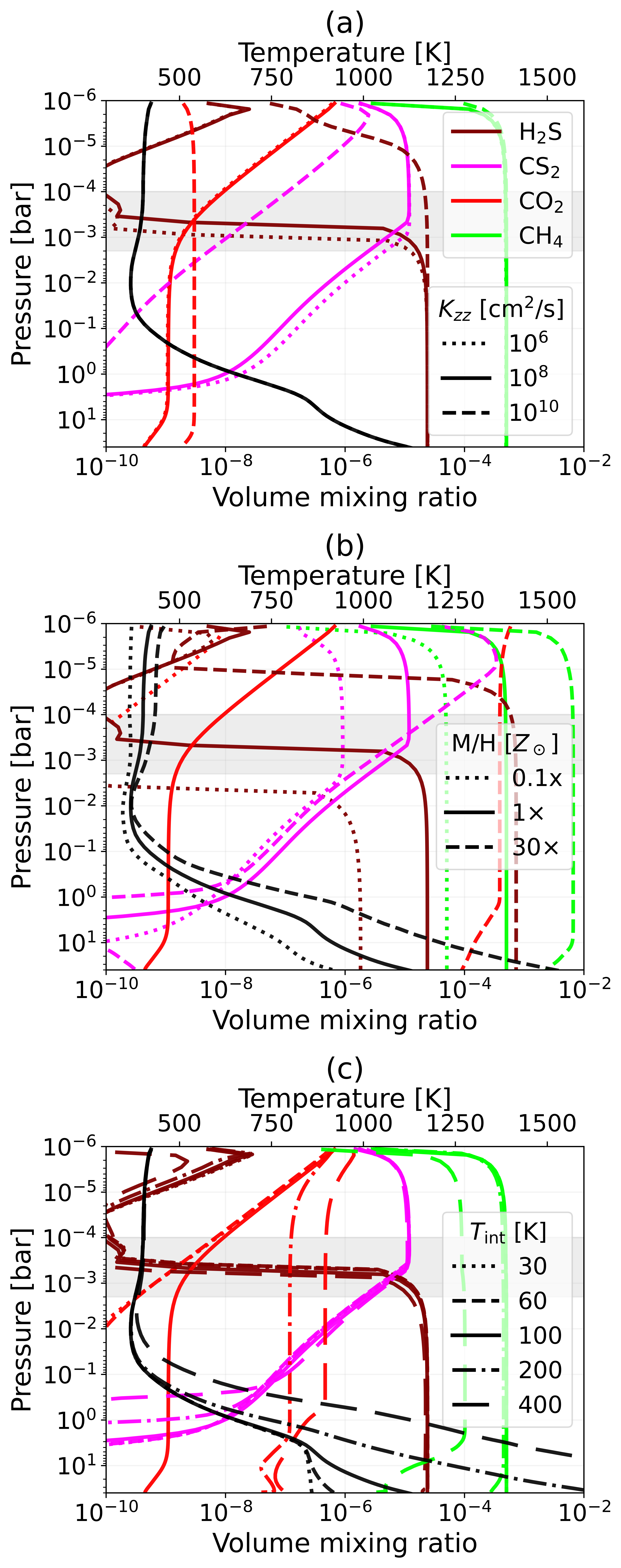}
     \caption{\footnotesize Sensitivity analysis of the atmosphere of TOI-6894~b for various model input parameters. The reference model assumes $K_{\rm zz}=10^8$ cm$^2$ s$^{-1}$, $1\times Z_{\odot}$ metallicity, and $T_{\rm int}=100$\,K. The model parameters were varied for (a) $K_{\rm zz}$, (b) metallicity, and (c) $T_{\rm int}$. Colors denote different chemical species, while line styles indicate the values adopted for the varied parameter in each panel. Solid lines correspond to the reference model. $T_{\rm int}$ has the smallest effect on \ce{CS2} formation among the parameters considered.}
    \label{fig:sensitivity_analysis}
\end{figure}

\subsection{Sensitivity Analysis of \ce{CS2} Formation to Atmospheric Parameters} \label{sec:sensitivity_analysis_atmospheric_parameters}

Figure~\ref{fig:sensitivity_analysis} shows the sensitivity of the modeled atmosphere of TOI-6894~b to various atmospheric parameters, including $K_{\rm zz}$, metallicity, and $T_{\rm int}$. As shown in Figure~\ref{fig:sensitivity_analysis}a, the atmospheric chemistry exhibits only minor variations for $K_{\rm zz}$ values up to $10^8$ cm$^2$ s$^{-1}$. However, when a higher vertical mixing coefficient of $K_{\rm zz}=10^{10}$ cm$^2$ s$^{-1}$ is assumed, the model predicts significantly lower \ce{CS2} abundances. This behavior arises because stronger vertical mixing reduces the relative contribution of photochemistry within the JWST-observable region (grey shaded region in Figure~\ref{fig:sensitivity_analysis}a), leading to a decrease in photochemically produced \ce{CS2} and an enhanced transport of \ce{H2S} from the deeper atmosphere. In other words, the region dominated by vertical mixing extends toward higher altitudes and replaces part of the upper atmosphere that is photochemically controlled under lower $K_{\rm zz}$ conditions. This nonlinear response indicates that $K_{\rm zz}\sim10^8$--$10^{10}$ cm$^2$ s$^{-1}$ represents a transition regime, where vertical mixing becomes sufficiently efficient to compete with photochemistry and substantially control the \ce{CS2} abundance.

As discussed in Sections~\ref{sec:cs2_formation_pathways} and \ref{sec:sensitivity_analysis_rate_coefficients}, this further supports our conclusion that \ce{CS2} is primarily a photochemical product in the atmosphere of TOI-6894~b and is preferentially produced in the upper atmosphere where photochemical processes dominate.

Varying the atmospheric metallicity primarily controls the maximum abundance of \ce{CS2} formed in the atmosphere, whereas the conversion of sulfur from \ce{H2S} to \ce{CS2} within the intermediate pressure region ($P\sim10^{-3}$--1 bar) exhibits only minor differences, as shown in Figure~\ref{fig:sensitivity_analysis}b. Increasing the metallicity enhances the total inventories of C, O, and S, leading to higher abundances of \ce{CH4}, \ce{CO2}, and \ce{H2S}. Consequently, the absolute abundance of \ce{CS2} increases with metallicity, while the photochemical conversion efficiency from \ce{H2S} to \ce{CS2} remains largely unchanged. This behavior likely occurs because the photochemistry is primarily controlled by available UV photons in the upper atmosphere rather than by the absolute sulfur inventory. These results suggest that sulfur in the atmosphere of TOI-6894~b can be efficiently converted into \ce{CS2} via photochemistry over the explored metallicity range from 0.1$\times$ to 30$\times Z_{\odot}$.

Varying $T_{\rm int}$ primarily affects the deep atmospheric chemical composition, particularly the abundances of \ce{CH4}, CO, and \ce{CO2}. The mixing ratios of these species are determined by their thermochemical stability in the deep atmosphere and subsequently quenched by vertical mixing, allowing them to remain abundant within the JWST-observable region \citep{Moses_2011, Fortney_2020, yang2024chemical}. As discussed in Section~\ref{sec:cs2_formation_pathways}, \ce{CH4} acts as the primary carbon source for \ce{CS2} formation through photochemistry. Therefore, the abundance of \ce{CS2} is expected to depend on the quenched abundance of deep atmospheric \ce{CH4}, and consequently on $T_{\rm int}$. However, as shown in Figure~\ref{fig:sensitivity_analysis}c, the modeled \ce{CS2} abundance exhibits only minor variations within the JWST-observable region over the explored range of $T_{\rm int}=30$--400\,K for the 1$\times Z_{\odot}$ atmosphere of TOI-6894~b. This result suggests that the quenched \ce{CH4} abundance above $\sim100$ ppmv (VMR $\gtrsim10^{-4}$) remains sufficiently high to sustain efficient photochemical \ce{CS2} production throughout the $T_{\rm int}$ range of 30--400\,K in the atmosphere of TOI-6894~b. For planets with higher $T_{\rm eq}$ and $T_{\rm int}$, however, thermochemical depletion of \ce{CH4} may eventually limit \ce{CS2} production, as discussed in Section~\ref{sec:teq_dependence_of_CS2_formation}.

Although not shown, varying the stellar spectrum from GJ~876 to the more UV-active AD Leo enhances \ce{CS2} production deeper ($>$1 mbar) in the atmosphere by increasing the availability of 100--400 nm photons and extending the region where photochemistry dominates, increasing \ce{CS2} abundances by up to an order of magnitude. However, the resulting changes near the $\sim$1 mbar level most relevant to JWST transmission spectroscopy are comparatively modest, with \ce{CS2} abundances increasing by less than a factor of three and producing almost no discernible change in the model transmission spectra (Section~\ref{sec:model_transmission_vs_JWST}). We therefore omit these results for brevity, as they do not alter our conclusions. Overall, these findings suggest that our conclusions regarding photochemical \ce{CS2} production are robust to plausible variations in stellar UV activity.

\begin{figure*}[htb!]
    \includegraphics[width=1\textwidth]{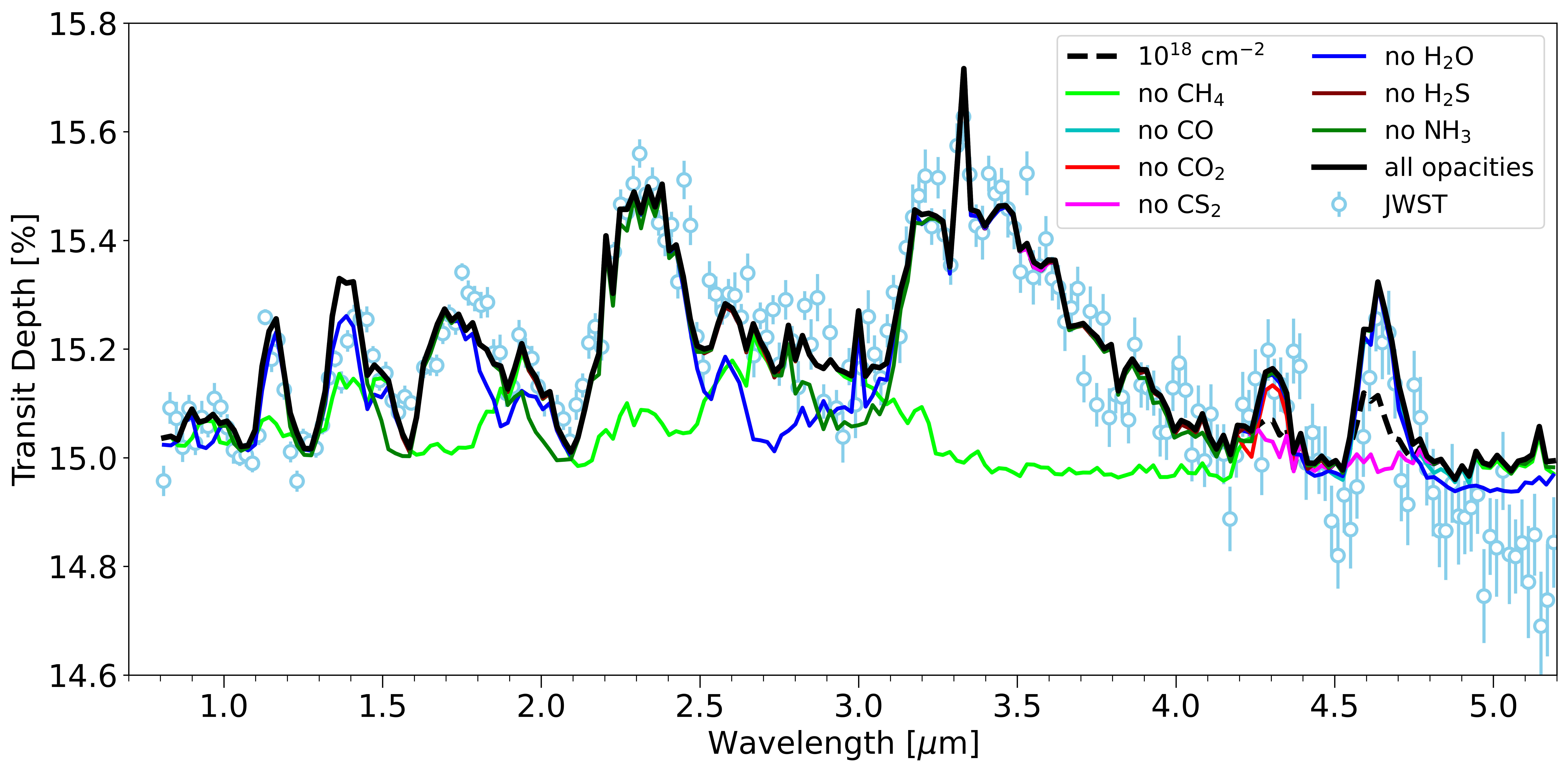}
     \caption{\footnotesize Model-predicted transmission spectrum based on the fiducial model (3$\times Z_{\odot}$, $K_{\rm zz}=10^8$ cm$^2$ s$^{-1}$, and $T_{\rm int}=100$\,K) presented in Figure~\ref{fig:3x_vmr}. The black solid line represents the simulated transmission spectrum including all opacity sources, while the black dashed line shows the same model with the \ce{CS2} abundance adjusted to yield a column density of 10$^{18}$ molecules cm$^{-2}$ within the observable atmosphere, corresponding to the approximate detectability threshold (refer to Section~\ref{sec:teq_dependence_of_CS2_formation}). Colored solid lines show simulated transmission spectra computed while excluding the opacity contribution from individual molecular species, as indicated. Light blue symbols with error bars denote the JWST observations presented in the companion observational study by \citet{zhang2026cnos}}
    \label{fig:transmission_spectra}
\end{figure*}

\subsection{Model-Predicted Transmission Spectra of TOI-6894~b Compared with JWST Observations} \label{sec:model_transmission_vs_JWST}

Figure~\ref{fig:transmission_spectra} compares the model-predicted transmission spectra generated with \texttt{PLATON} \citep{Zhang_2019_platon, Zhang_2025_platon}, based on the fiducial model (3$\times Z_{\odot}$, $K_{\rm zz}=10^8$ cm$^2$ s$^{-1}$, and $T_{\rm int}=100$\,K), with JWST observations of TOI-6894~b \citep{zhang2026cnos}. The predicted spectra are broadly consistent with the NIRSpec/PRISM observation and reproduce the absorption features of \ce{CH4} (lime), \ce{NH3} (green), \ce{H2O} (blue), and \ce{CS2} (magenta).

Notably, the feature near 4.2~$\mu$m, which is often attributed to \ce{CO2} absorption in exoplanet atmospheres \citep[e.g., WASP-39~b;][]{jtec2023}, is instead attributed here to the \ce{CS2} band $2\nu_{2}$~(bending)~+~$\nu_{3}$~(asymmetric stretching), corresponding to $\sim$2304~cm$^{-1}$ \citep{plyler1947infrared}. This feature may therefore help constrain the atmospheric metallicity of TOI-6894~b, because metallicities higher than $3\times Z_{\odot}$ would produce additional \ce{CO2} absorption in this wavelength region, increasing the discrepancy between the model and the observations.

The 4.6~$\mu$m feature is more clearly attributable to \ce{CS2}. However, at higher metallicity, CO absorption may partially mask this region, making it more difficult to distinguish between \ce{CS2} and CO. This degeneracy may help explain the case of V1298~Tau~b \citep{barat2025metal}, which will be discussed in Section~\ref{sec:teq_dependence_of_CS2_formation}, and may also account for the tentative feature reported in the atmosphere of TOI-270~d.

\begin{figure*}[htb!]
    \centering    
    \includegraphics[width=1\textwidth]{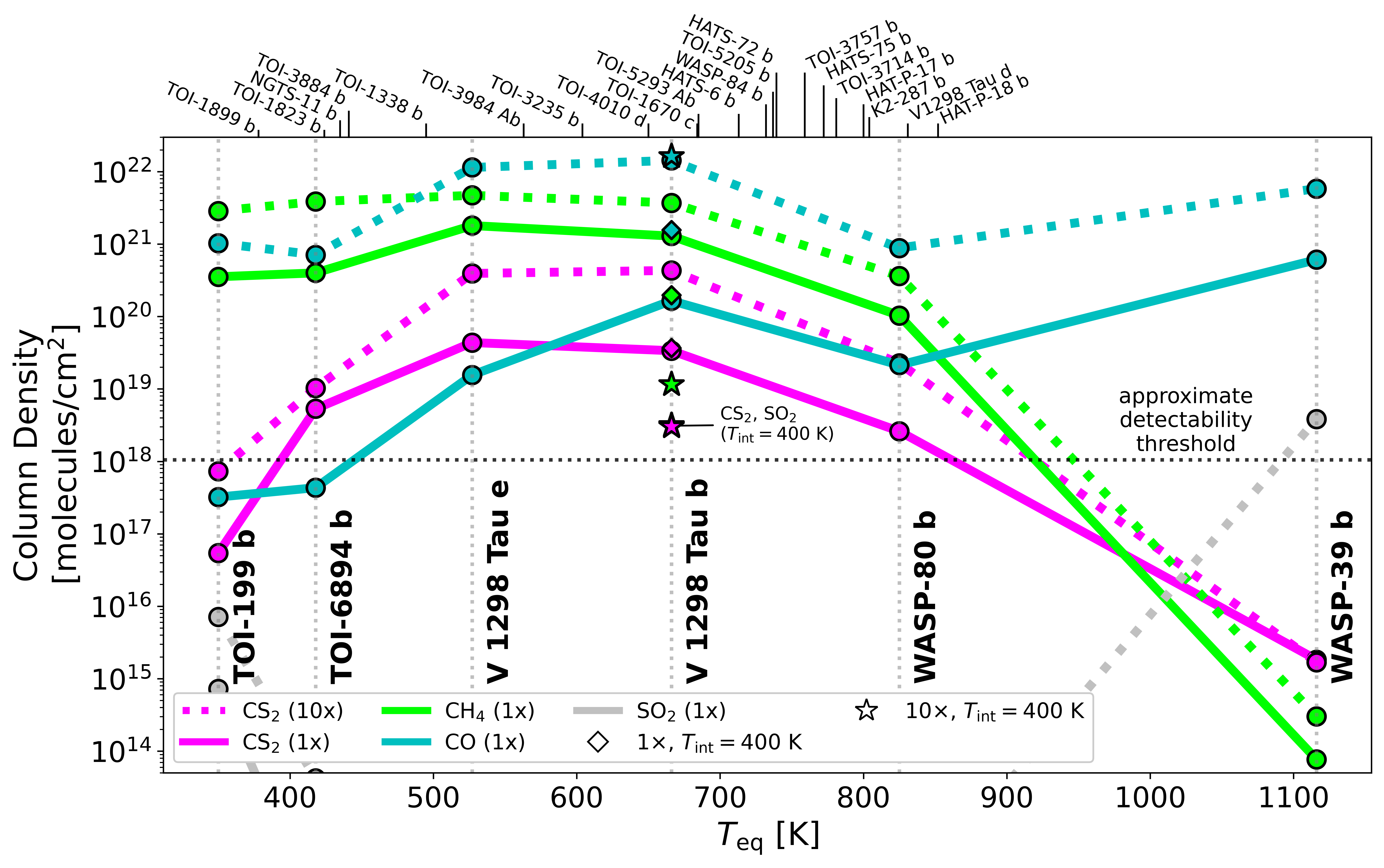}
     \caption{\footnotesize Dependence of the predicted \ce{CS2}, \ce{CH4}, CO, and \ce{SO2} abundances, quantified as the column density (molecules cm$^{-2}$) integrated over the \mbox{JWST}-observable pressure region ($P\leq2$ mbar), across a sample of gas giant exoplanets spanning $T_{\rm eq}=350$--1120\,K, assuming $T_{\rm int}=100$\,K and $K_{\rm zz}=10^8$ cm$^2$ s$^{-1}$ for both 1$\times$ and 10$\times Z_{\odot}$. Colors denote different chemical species, while line styles indicate metallicity: solid lines represent 1$\times Z_{\odot}$ and dotted lines represent 10$\times Z_{\odot}$. The predicted \ce{SO2} column densities remain below 10$^{16}$ molecules cm$^{-2}$ for all planets considered here, except for the 10$\times Z_{\odot}$ model of WASP-39~b. For V1298~Tau~b, additional models with $T_{\rm int}=400$\,K were computed at both 1$\times$ and 10$\times Z_{\odot}$ and are denoted by diamond and star symbols, respectively. The predicted \ce{CS2} and \ce{SO2} column densities for the $T_{\rm int}=400$ K models are nearly identical and therefore overlap. The black horizontal dotted line indicates the estimated \ce{CS2} detectability threshold column density of 10$^{18}$ molecules cm$^{-2}$. This corresponds approximately to a volume mixing ratio of 1 ppmv ($10^{-6}$). Planets labeled along the top x-axis are gas giants ($R\geq0.50 \,R_J$, $M\leq1.1\,M_J$) that have been or are planned to be observed with JWST using modes capable of probing \ce{CS2} absorption (i.e., NIRSpec/G395M, G395H, or PRISM). Their corresponding $T_{\rm eq}$ values, ranging from 350 to 852 K, are also indicated. These targets represent potential candidates for future \ce{CS2} detections.}
    \label{fig:teq_column_density}
\end{figure*}

\subsection{\ce{CS2} Formation across Gas Giant Exoplanet Atmospheres with Different $T_{\rm eq}$} \label{sec:teq_dependence_of_CS2_formation}

Figure~\ref{fig:teq_column_density} shows the column densities of CO, \ce{CH4}, \ce{SO2}, and \ce{CS2} integrated over the JWST-observable pressure region ($P\leq2$ mbar) as a function of $T_{\rm eq}$. As expected, \ce{CH4} and CO exhibit opposite overall trends with increasing $T_{\rm eq}$: \ce{CH4} decreases, whereas CO increases. This behavior reflects the thermochemical stability transition between \ce{CH4} and CO \citep{Moses_2011}, although some scatter is seen in the CO-$T_{\rm eq}$ trend line because the planets differ in parameters other than $T_{\rm eq}$. In contrast to \ce{CH4}, the CO column density increases substantially from 1$\times Z_{\odot}$ to 10$\times Z_{\odot}$, by nearly three orders of magnitude rather than the approximately order-of-magnitude increase seen for \ce{CH4}. At 10$\times Z_{\odot}$, the CO column density becomes sufficiently large that its dependence on $T_{\rm eq}$ is comparatively weak.

Because \ce{CH4} serves as the primary carbon source for \ce{CS2} formation, one might expect the \ce{CS2} abundance to follow a similar trend. However, the predicted \ce{CS2} column density instead exhibits a peaked distribution, reaching a maximum near $T_{\rm eq}\sim500$--700\,K. This behavior arises because efficient \ce{CS2} production requires both active photochemistry and sufficient thermal energy to sustain the underlying chemical pathways linking \ce{CH4} and \ce{H2S} to \ce{CS2}. While photochemistry in the upper atmosphere is essential for liberating sulfur atoms from \ce{S2}, adequate thermal energy is likewise required to drive the thermochemical reactions that liberate methyl radicals (\ce{CH3}) and sulfur atoms from initial \ce{CH4} and \ce{H2S}, respectively. At lower temperatures, such as for TOI-199~b ($T_{\rm eq}=350$\,K), the atmosphere lacks sufficient thermal energy for efficient \ce{CS2} production despite the high abundance of \ce{CH4}. As $T_{\rm eq}$ increases, thermochemical conversion becomes more efficient, leading to enhanced \ce{CS2} production that peaks at intermediate $T_{\rm eq}$. At yet higher $T_{\rm eq}$, however, the decreasing abundance of \ce{CH4} now limits the available carbon source for \ce{CS2} formation, resulting in declining \ce{CS2} column densities.

The potential detectability of \ce{CS2} can be estimated empirically from our transmission spectrum calculations. By varying the atmospheric \ce{CS2} abundance in our fiducial model, we find that a \ce{CS2} column density of $\gtrsim10^{18}$ molecules cm$^{-2}$ within the observable atmosphere is required to generate a spectral feature with an amplitude comparable to the current JWST measurement uncertainties for TOI-6894~b ($\sim1\sigma$), as shown by black dashed line in Figure~\ref{fig:transmission_spectra}. In terms of abundance, this roughly corresponds to a volume mixing ratio of $\sim$1 ppmv. From this perspective, the absence of \ce{CS2} feature in the JWST observation of TOI-199~b \citep{Bello-Arufe_2026} is consistent with the low predicted \ce{CS2} column densities in both the 1$\times$ and 10$\times Z_{\odot}$ cases.

In contrast, both TOI-6894~b and V1298~Tau~e exhibit substantial predicted \ce{CS2} column densities (see Figure~\ref{fig:teq_column_density}). More importantly, for the 1$\times Z_{\odot}$ scenarios, the predicted \ce{CS2} column densities exceed those of CO in both atmospheres. This distinction is particularly important for observational detectability because the major absorption bands of CO and \ce{CS2} overlap near $\sim$4.6~$\mu$m. Although the CO absorption lines are relatively sparse in this wavelength region, their absorption cross sections are approximately an order of magnitude larger than those of \ce{CS2}. Consequently, a higher \ce{CS2} column density relative to CO is favorable for distinguishing \ce{CS2} absorption from overlapping CO features. This result is consistent with the detections of \ce{CS2} in both TOI-6894~b and V1298~Tau~e \citep{zhang2026cnos,dai2026photochemicalcs2gasdetected}. However, the attenuated \ce{CH4} features and enhanced \ce{CO2} absorption reported for V1298~Tau~e by \citet{dai2026photochemicalcs2gasdetected} may indicate a lower C/O ratio than the solar value of 0.55 \citep{lodders2021relative}, $T_{\rm int}\geq100$\,K, or a combination of both.

V1298~Tau~b represents the opposite regime, where the CO column density exceeds that of \ce{CS2}. Although its $T_{\rm eq}\sim$670 K lies within the regime where our models predict potentially observable \ce{CS2}, recent JWST observations revealed prominent CO and \ce{CO2} absorption together with weaker \ce{CH4} features and a tentative detection of \ce{SO2} \citep{barat2025metal}. The enhanced CO abundance may partially obscure the 4.6~$\mu$m \ce{CS2} feature, but this alone is unlikely to fully explain the apparent absence of \ce{CS2}, particularly given the reported detection of \ce{CS2} in the hotter planet WASP-80~b. Instead, V1298~Tau~b may illustrate the importance of additional atmospheric parameters beyond $T_{\rm eq}$. As a young giant planet, V1298~Tau~b is expected to possess a substantial internal heat flux, and atmospheric retrievals suggest either a high internal temperature ($T_{\rm int}\lesssim500$ K) and/or a sub-solar C/O ratio of 0.22 \citep{barat2025metal}. Both effects reduce atmospheric \ce{CH4} abundances, thereby suppressing \ce{CS2} production in our models. Consistent with this interpretation, our $T_{\rm int}=400$\,K models predict lower \ce{CH4} abundances, significantly enhanced CO, comparable \ce{SO2} and \ce{CS2} column densities (both close to the estimated detectability threshold), and reduced \ce{CS2}/CO contrast compared to the $T_{\rm int}=100$\,K cases. Given that our $T_{\rm int}=400$\,K models qualitatively reproduce the key spectroscopic characteristics of V1298~Tau~b, we therefore speculate that the apparent absence of \ce{CS2} in V1298~Tau~b reflects the combined influence of internal heating, atmospheric composition, and spectral masking by CO-bearing species, rather than $T_{\rm eq}$ alone.

At the $T_{\rm eq}$ of WASP-80~b ($\sim$830\,K), the atmosphere enters a regime where the abundances of CO and \ce{CH4} become comparable. Recent JWST/NIRCam and MIRI observations of WASP-80~b identified a prominent absorption feature near 4.6 and 6.5~$\mu$m that may be attributed to \ce{CS2} \citep{triantafillides2026identification}. The relative column density ratios among \ce{CH4}, CO, and \ce{CS2} in WASP-80~b are similar to those predicted for V1298~Tau~b, although the absolute column densities are approximately an order of magnitude lower (Figure~\ref{fig:teq_column_density}). From our modeling perspective, the predicted \ce{CS2} column density in WASP-80~b remains above $10^{18}$ [molecules cm$^{-2}$] even for the 1$\times Z_{\odot}$ scenario, while the 10$\times Z_{\odot}$ model predicts an additional order-of-magnitude enhancement in \ce{CS2} abundance. However, the CO column density is also predicted to increase substantially, exceeding that of \ce{CS2} by approximately a factor of 10 for the 1$\times Z_{\odot}$ scenario and by nearly two orders of magnitude for the 10$\times Z_{\odot}$ scenario (Figure~\ref{fig:teq_column_density}). These results suggest that the atmosphere of WASP-80~b can maintain a detectable abundance of \ce{CS2}. However, the presence of substantial amounts of CO, despite its relatively sparse absorption lines in this wavelength region, may complicate the distinction between \ce{CS2} and CO features near 4.6~$\mu$m. This interpretation is broadly consistent with the JWST observations reported by \citet{triantafillides2026identification}.

At the $T_{\rm eq}$ of WASP-39~b (1120\,K), \ce{CH4} is completely depleted from the atmosphere, with the carbon inventory primarily locked in CO. Thus, \ce{CS2} formation becomes limited by the lack of available \ce{CH4}, leading to the lowest \ce{CS2} column densities (see Figure~\ref{fig:teq_column_density}). In contrast, \ce{SO2} becomes abundant at 10$\times Z_{\odot}$. These predictions are consistent with the extensive JWST observations of WASP-39~b, which have revealed prominent \ce{CO2} and \ce{SO2} features with no \ce{CH4} and \ce{CS2} features \citep{jtec2023, Rustamkulov-2023, Ahrer-2023, Powell_2024}. Together with the \ce{CS2} detections in WASP-80~b, these results suggest the existence of a transition $T_{\rm eq}$, between 830 and 1120\,K, at which the dominant observable sulfur-bearing species shifts from \ce{CS2} to \ce{SO2}.

Overall, our photochemical modeling spanning a wide range of $T_{\rm eq}$ suggests that detectable \ce{CS2} abundances are favored in gas giants with $T_{\rm eq}\sim400$--800\,K. A number of additional planets within this $T_{\rm eq}$ range have already been or are planned to be observed with JWST using modes capable of probing \ce{CS2} absorption (e.g., NIRSpec/G395M, G395H, and PRISM), making them promising targets for future \ce{CS2} detections (see labels along the top $x$-axis of Figure~\ref{fig:teq_column_density}). 

Furthermore, recent studies have shown that \ce{SO2} is highly sensitive to atmospheric metallicity and serves as a powerful tracer of heavy-element enrichment in warm and hot giant exoplanets \citep{Tsai_2023, Crossfield_2023, Powell_2024, Crossfield_2025_SO2_shorline}. Our results suggest that \ce{CS2} may play an analogous role in cooler gas giants, where \ce{SO2} becomes less abundant. Although the \ce{CS2} abundance can be influenced by factors such as sub-solar C/O ratios, elevated $T_{\rm int}$, and stellar UV activity, its strong dependence on metallicity makes it a promising sulfur-based metallicity tracer. Together, \ce{CS2} and \ce{SO2} provide complementary probes of atmospheric sulfur chemistry and metallicity across a wide range of gas giants.

\section{Conclusions} \label{sec:conclusions}
In this Letter, we performed one-dimensional radiative-convective and photochemical-kinetic transport atmospheric modeling of TOI-6894~b, along with an extensive sensitivity analysis to investigate the photochemical production of \ce{CS2}. Our results are broadly consistent with the JWST NIRSpec/PRISM observations and support the interpretation that the observed 4.6 $\mu$m feature arises from photochemically produced \ce{CS2}. The formation of \ce{CH4} and \ce{H2S} in the deep atmosphere is essential for subsequent \ce{CS2} production, since these species serve as the primary carbon and sulfur reservoirs. Through the combined effects of thermochemical and photochemical processes, carbon and sulfur initially stored in \ce{CH4} and \ce{H2S} are progressively converted into \ce{H2CS}, HCS, CS, and ultimately \ce{CS2}. Our sensitivity analysis identifies \ce{S2} photolysis as a key process driving the disequilibrium chemistry that leads to \ce{CS2} formation in TOI-6894~b. 

By extending our analysis to a broader $T_{\rm eq}=350$--1120\,K, we find that detectable \ce{CS2} abundances are favored in temperate-to-warm gas-giant atmospheres ($T_{\rm eq}\sim400$--800\,K). This $T_{\rm eq}$ range encompasses 17 gas giant exoplanets that have already been or are planned to be observed with JWST using observing modes sensitive to \ce{CS2}, making them promising targets for future detections (Figure~\ref{fig:teq_column_density}). In particular, our models predict substantial \ce{CS2} production in TOI-6894~b,  V1298~Tau~e, and WASP-80~b, consistent with the recent detection of \ce{CS2} in these planets \citep{triantafillides2026identification, dai2026photochemicalcs2gasdetected, zhang2026cnos}, as well as the absence of \ce{CS2} in V1298~Tau~b \citep{barat2025metal}. In summary, our modeling results suggest that \ce{CS2} can serve as a valuable tracer of the sulfur inventory in temperate to warm gas giants ($T_{\rm eq}\sim400$ -- 800\,K), complementing \ce{SO2}, which is expected to be a more effective sulfur tracer in hotter giant exoplanets \citep{Tsai_2023, Crossfield_2023, Powell_2024, Crossfield_2025_SO2_shorline}.

\begin{acknowledgments}
The authors gratefully acknowledge Dr. Alexander W. Hull and Professor Robert W. Field at Massachusetts Institute of Technology for insightful discussions on \ce{S2} photolysis. The authors also thank Dr. Saugata Barat at Massachusetts Institute of Technology and Dr. Sagnick Mukherjee at Arizona State University for providing the $T$--$P$ profiles for V1298~Tau~b. V.N. acknowledges support from the National Science Foundation Graduate Research Fellowship under Grant No. DGE 2140001. C.P.-G. acknowledges support from the E. Margaret Burbidge Prize Postdoctoral Fellowship from the Brinson Foundation, and the Suzuki Fellowship from the Yuji and Lorraine Suzuki Postdoctoral Research Fund.  
This work is based on observations made with the NASA/ESA/CSA James Webb Space Telescope. The data were obtained from the Mikulski Archive for Space Telescopes at the Space Telescope Science Institute, which is operated by the Association of Universities for Research in Astronomy, Inc., under NASA contract NAS 5-03127 for JWST. These observations are associated with program \#GO-8696. Support for this program was provided by NASA through a grant from the Space Telescope Science Institute.
\end{acknowledgments}

\begin{contribution}

J.Y. conceptualized and led the project. J.Y., M.Z, E.M.-R.K., and J.L.B. jointly designed the study. J.Y. led the writing of the manuscript. J.Y., V.N., and M.R.L. conducted 1D photochemical kinetic-transport modeling. J.Y., V.N., and M.R.L. conducted the 1D-RCE calculations. J.Y. conducted the sensitivity analysis and generated the model transmission spectra of TOI-6894 b. Q.X. and G.F. conducted JWST data reduction. J.Y., V.N., M.Z., Q.X, E.M.-R.K., J.L.B., J.J.F., P.G., M.C.N., C.P.-G., K.B.S., M.B., J.P.W., and L.W. contributed to the writing the manuscript. M.Z., J.L.B., M.R.L., J.J.F., P.G., M.C.N., C.P.-G., K.B.S., M.B., J.P.W., L.W., G.F., J.M.D., V.P., and D.P are on the original JWST proposal (\#GO-8696)


\end{contribution}

%

\software{ \texttt{PICASO 4.0} \citep{mang2026picaso}, \texttt{EPACRIS} \citep{yang2024epacris}, \texttt{PLATON} \citep{Zhang_2019_platon, Zhang_2025_platon}}


\appendix

\section{Chemical Network Details} \label{sec:appendix_chemical_network}

The photochemical network was adopted from \citet{yang2024chemical}, where it was constructed using the Reaction Mechanism Generator \citep[\texttt{RMG};][]{Gao_2016, rmg-v3, RMG-database}, a rate-based automated chemical network generation framework widely used in chemical engineering \citep{Liu_2020, Dong-2023}. Built upon a continuously maintained database of experimental measurements and high-level \textit{ab-initio} calculations for thermodynamic and kinetic parameters \citep{RMG-database}, \texttt{RMG} systematically explores chemical reaction networks and identifies chemically relevant species and reactions through an automated, flux-based algorithm. 

Using \texttt{RMG} enables the construction of chemically comprehensive and self-consistent reaction mechanisms while minimizing the risk of omitting important species and reactions. At the same time, species and reactions that contribute negligibly to a system of interest are excluded, reducing computational expense without sacrificing chemical fidelity. A previous benchmarking study of Jupiter's \ce{CO}-\ce{CH4} chemistry demonstrated that \texttt{RMG}-generated networks are more robust to uncertainties in individual reaction rates than other networks, owing to \texttt{RMG}'s comprehensive and systematic treatment of the reaction system \citep{yang2026jupiter}.

A current limitation of \texttt{RMG}, however, is that photochemistry is not considered during network generation. Consequently, species that are produced primarily through photochemistry are normally omitted from the resulting mechanism despite their important roles in atmospheric photochemistry. This is the case for \ce{CS2}, which is generally not expected to be abundant under thermochemical equilibrium. 

To account for this limitation and incorporate \ce{CS2} chemistry, we augmented the \texttt{RMG}-generated network with all reduced sulfur-bearing species and their associated reactions that were absent from the \texttt{RMG} database but available in the \texttt{VULCAN} chemical network \citep{Tsai_2017,Tsai_2021} as described in Section~\ref{subsec:chemical_network}. The resulting chemical network contains 103 species, including 33 sulfur-bearing species, and 2067 reactions.

We additionally tested two alternative chemical networks: the full \texttt{VULCAN} \texttt{SNCHO\_photo\_network\_2025} \citep{VULCAN_SNCHO_2025}, consisting of 91 species (21 sulfur-bearing species) and 593 reactions, and the \texttt{Photochem} network \citep{wogan2025photochem}, containing 92 species (20 sulfur-bearing species) and 682 reactions. Although not shown in this Letter, only the \texttt{Photochem} network failed to produce sufficient \ce{CS2} within the JWST-observable region \citep[$P\sim0.1{-}2$ mbar;][]{Rustamkulov-2023}. This difference is briefly discussed further in Section~\ref{sec:cs2_formation_pathways}.

\section{1D Photochemical Modeling Setup for Various Gas Giant Exoplanets Other Than TOI-6894~b} \label{sec:other_exoplanets}
We additionally explored five gas giant exoplanets that have been observed by the JWST/NIRSpec instrument and that span a range of $T_{\rm eq}$  relevant for \ce{CS2} formation:

\noindent TOI-199~b \citep[$T_{\rm eq}\sim$ 350\,K;][]{Bello-Arufe_2026}, V1298~Tau~e \citep[$T_{\rm eq}\sim$530\,K;][]{livingston2026young}, V1298~Tau~b \citep[$T_{\rm eq}\sim$670\,K;][]{livingston2026young}, WASP-80~b \citep[$T_{\rm eq}\sim$830\,K;][]{triaud-2015}, and WASP-39~b \citep[$T_{\rm eq}\sim$ 1120\,K;][]{mancini2018gaps}.

To consistently investigate the $T_{\rm eq}$ dependence of \ce{CS2} formation, we fixed $K_{zz}=10^8$ cm$^2$ s$^{-1}$ and $T_{\rm int}=100$\,K, and adopted the photochemical network described in Section~\ref{subsec:epacris} for all atmospheric simulations. Thus, the major differences from the TOI-6894~b case are the adopted $T$--$P$ profiles and stellar fluxes used in the 1D photochemical modeling.
\begin{description}[topsep=5pt]
\item[TOI-199~b]
We adopted the $T$--$P$ profiles assuming $T_{\rm int}=100$\,K for both 1$\times$ and 10$\times Z_{\odot}$ cases, and used the stellar spectrum of TOI-199 from \citet{Bello-Arufe_2026}. 
\item[V1298~Tau~b and e]
We computed $T$--$P$ profiles assuming $T_{\rm int}=100$\,K for both 1$\times$ and 10$\times Z_{\odot}$ cases using \texttt{PICASO} assuming full day--night heat redistribution, as described in Section~\ref{subsec:picaso}.
We additionally computed profiles assuming $T_{\rm int}=400$\,K for both metallicities to further explore the dependence on interior heating, given that this is a young system. For the stellar spectrum, we adopted the HD~97658 spectrum \citep[K1V type, from the \texttt{MUSCLES} Treasury Survey;][]{loyd_2016}, since V1298~Tau is also a K1-type star \citep{suarez2022rapid}. We then scaled this spectrum to reproduce the bolometric insolation received by V1298~Tau~b (35 $S_{\oplus}$; \citealt{David_2019_v1298tau_system}) and V1298~Tau~e (10 $S_{\oplus}$; \citealt{David_2019_v1298tau_system}). We note that V1298~Tau is a young star \citep[23$\pm$4 Myr;][]{david2019warm} and is therefore likely to emit stronger 100--400 nm UV flux than HD~97658. As discussed in Section~\ref{sec:sensitivity_analysis_atmospheric_parameters}, higher UV fluxes enhance \ce{CS2} production in our models. Therefore, the \ce{CS2} abundances predicted here for V1298~Tau~b and e should be regarded as conservative lower limits.
\item[WASP-80~b]
We adopted the $T$--$P$ profile from \citet{Bell-2023}, which assumes $T_{\rm int}=100$\,K and 10$\times Z_{\odot}$, for modeling both the 1$\times$ and 10$\times Z_{\odot}$ cases, and used a stellar flux of HD~85512 \citep[K6V;][]{loyd_2016} as in \citet{yang2024epacris}.
\item[WASP-39~b]
We adopted the morning-limb $T$--$P$ profile from \citet{Tsai_2023}, which assumes $T_{\rm int}=100$\,K and 10$\times Z_{\odot}$, for modeling both the 1$\times$ and 10$\times Z_{\odot}$ cases, and adopted the reconstructed stellar spectral energy distribution from \citet{Tsai_2023}, which combines HST/STIS observations of WASP-39 with proxy NUV and FUV/XUV spectra from HD~203244 and the quiet Sun, respectively.
\end{description}


\bibliography{references}{}

@ARTICLE{fu2024_h2s,
       author = {{Fu}, Guangwei and {Welbanks}, Luis and {Deming}, Drake and {Inglis}, Julie and {Zhang}, Michael and {Lothringer}, Joshua and {Ih}, Jegug and {Moses}, Julianne I. and {Schlawin}, Everett and {Knutson}, Heather A. and {Henry}, Gregory and {Greene}, Thomas and {Sing}, David K. and {Savel}, Arjun B. and {Kempton}, Eliza M.-R. and {Louie}, Dana R. and {Line}, Michael and {Nixon}, Matt},
        title = "{Hydrogen sulfide and metal-enriched atmosphere for a Jupiter-mass exoplanet}",
      journal = {\nat},
         year = 2024,
        month = aug,
       volume = {632},
       number = {8026},
        pages = {752-756},
          doi = {10.1038/s41586-024-07760-y},
archivePrefix = {arXiv},
       eprint = {2407.06163},
 primaryClass = {astro-ph.EP},
       adsurl = {https://ui.adsabs.harvard.edu/abs/2024Natur.632..752F}
}

@ARTICLE{Tsai_2023,
       author = {Tsai, Shang-Min and Lee, Elspeth K. H. and Powell, Diana and Gao, Peter and Zhang, Xi and Moses, Julianne and Hébrard, Eric and Venot, Olivia and Parmentier, Vivien and Jordan, Sean and Hu, Renyu and Alam, Munazza K. and Alderson, Lili and Batalha, Natalie M. and Bean, Jacob L. and Benneke, Björn and Bierson, Carver J. and Brady, Ryan P. and Carone, Ludmila and Carter, Aarynn L. and Chubb, Katy L. and Inglis, Julie and Leconte, Jérémy and Lopez-Morales, Mercedes and Miguel, Yamila and Molaverdikhani, Karan and Rustamkulov, Zafar and Sing, David K. and Stevenson, Kevin B. and Wakeford, Hannah R and Yang, Jeehyun and Aggarwal, Keshav and Baeyens, Robin and Barat, Saugata and Borro, Miguel de Val and Daylan, Tansu and Fortney, Jonathan J. and France, Kevin and Goyal, Jayesh M and Grant, David and Kirk, James and Kreidberg, Laura and Louca, Amy and Moran, Sarah E. and Mukherjee, Sagnick and Nasedkin, Evert and Ohno, Kazumasa and Rackham, Benjamin V. and Redfield, Seth and Taylor, Jake and Tremblin, Pascal and Visscher, Channon and Wallack, Nicole L. and Welbanks, Luis and Youngblood, Allison and Ahrer, Eva-Maria and Batalha, Natasha E. and Behr, Patrick and Berta-Thompson, Zachory K. and Blecic, Jasmina and Casewell, S. L. and Crossfield, Ian J. M. and Crouzet, Nicolas and Cubillos, Patricio E. and Decin, Leen and Désert, Jean-Michel and Feinstein, Adina D. and Gibson, Neale P. and Harrington, Joseph and Heng, Keivn and Henning, Thomas and Kempton, Eliza M. -R. and Krick, Jessica and Lagage, Pierre-Olivier and Lendl, Monika and Line, Michael and Lothringer, Joshua D. and Mansfield, Megan and Mayne, N. J. and Mikal-Evans, Thomas and Palle, Enric and Schlawin, Everett and Shorttle, Oliver and Wheatley, Peter J. and Yurchenko, Sergei N.},
        title = "{Photochemically produced \ce{SO2} in the atmosphere of WASP-39b}",
      journal = {Nature},
         year = "2023",
        month = "Apr",
       volume = {617},
        pages ={483--487},
          doi = {10.1038/s41586-023-05902-2}
}

@ARTICLE{Moses_2011,
    author = {{Moses}, J. I. and {Visscher}, C. and {Fortney}, J. J. and {Showman}, A. P. and {Lewis}, N. K. and {Griffith}, C. A. and {Klippenstein}, S J. and {Shabram}, M. and {Friedson}, A. J. and {Marley}, M. S. and {Freedman}, R. S.},
        title = "{Disequilibrium Carbon, Oxygen, and Nitrogen Chemistry in the Atmospheres of HD 189733b and HD 209458b}",
      journal = {\apj},
         year = "2011",
       volume = 737,
       number = 1,
          doi = {https://doi.org/10.1088/0004-637X/737/1/15},
}

@ARTICLE{Tsai_2017,
       author = {Tsai, Shang-Min and Lyons, James R. and Grosheintz, Luc and Rimmer, Paul B. and Kitzmann, Daniel and Heng, Kevin},
        title = "{VULCAN: an Open-Source, Validated Chemical Kinetics Python Code for Exoplanetary Atmospheres}",
      journal = {Astrophys. J. Suppl. Ser.},
         year = "2017",
       volume = {228},
       number = {2},
          doi = {10.3847/1538-4365/228/2/20}
}

@ARTICLE{Gao_2016,
       author = {{Gao}, Connie W. and {Allen}, Joshua W. and {Green}, William H. and {West}, Richard H.},
        title = "{Reaction Mechanism Generator: Automatic construction of chemical kinetic mechanisms}",
      journal = {Comput. Phys. Commun.},
         year = "2016",
        volume = 203,
        pages = {212-225},
          doi = {10.1016/j.cpc.2016.02.013},
}

@ARTICLE{Liu_2020,
       author = {{Liu}, Mengjie and {Chu}, Te-Chun and {Jocher}, Agnes and {Smith}, Mica C. and {Lengyel}, Istvan and {Green}, William H.},
        title = "{Predicting polycyclic aromatic hydrocarbon formation with an automatically generated mechanism for acetylene pyrolysis}",
      journal = {Int. J. Chem. Kinet.},
         year = "2020",
        month = "Aug",
        volume = 53,
        issue = 1,
        pages = {27-42},
          doi = {10.1002/kin.21421},
}

@article{Tsai_2021,
doi = {10.3847/1538-4357/ac29bc},
url = {https://dx.doi.org/10.3847/1538-4357/ac29bc},
year = {2021},
month = {dec},
publisher = {The American Astronomical Society},
volume = {923},
number = {2},
pages = {264},
author = {Shang-Min Tsai and Matej Malik and Daniel Kitzmann and James R. Lyons and Alexander Fateev and Elspeth Lee and Kevin Heng},
title = {A Comparative Study of Atmospheric Chemistry with VULCAN},
journal = {\apj}
}

@article{rmg-v3,
author = {Liu, Mengjie and Grinberg Dana, Alon and Johnson, Matthew S. and Goldman, Mark J. and Jocher, Agnes and Payne, A. Mark and Grambow, Colin A. and Han, Kehang and Yee, Nathan W. and Mazeau, Emily J. and Blondal, Katrin and West, Richard H. and Goldsmith, C. Franklin and Green, William H.},
title = {Reaction Mechanism Generator v3.0: Advances in Automatic Mechanism Generation},
journal = {Journal of Chemical Information and Modeling},
volume = {61},
number = {6},
pages = {2686-2696},
year = {2021},
doi = {10.1021/acs.jcim.0c01480},
    note ={PMID: 34048230},
URL = {https://doi.org/10.1021/acs.jcim.0c01480},
eprint = {https://doi.org/10.1021/acs.jcim.0c01480}

}

@ARTICLE{jtec2023,
       author = {{JWST Early Release Science Team} and {Ahrer}, Eva-Maria and {Alderson}, Lili and {Batalha}, Natalie M. and {Batalha}, Natasha E. and {Bean}, Jacob L. and {Beatty}, Thomas G. and {Bell}, Taylor J. and {Benneke}, Bj{\"o}rn and {Berta-Thompson}, Zachory K. and {Carter}, Aarynn L. and {Crossfield}, Ian J.~M. and {Espinoza}, N{\'e}stor and {Feinstein}, Adina D. and {Fortney}, Jonathan J. and {Gibson}, Neale P. and {Goyal}, Jayesh M. and {Kempton}, Eliza M. -R. and {Kirk}, James and {Kreidberg}, Laura and {L{\'o}pez-Morales}, Mercedes and {Line}, Michael R. and {Lothringer}, Joshua D. and {Moran}, Sarah E. and {Mukherjee}, Sagnick and {Ohno}, Kazumasa and {Parmentier}, Vivien and {Piaulet}, Caroline and {Rustamkulov}, Zafar and {Schlawin}, Everett and {Sing}, David K. and {Stevenson}, Kevin B. and {Wakeford}, Hannah R. and {Allen}, Natalie H. and {Birkmann}, Stephan M. and {Brande}, Jonathan and {Crouzet}, Nicolas and {Cubillos}, Patricio E. and {Damiano}, Mario and {D{\'e}sert}, Jean-Michel and {Gao}, Peter and {Harrington}, Joseph and {Hu}, Renyu and {Kendrew}, Sarah and {Knutson}, Heather A. and {Lagage}, Pierre-Olivier and {Leconte}, J{\'e}r{\'e}my and {Lendl}, Monika and {MacDonald}, Ryan J. and {May}, E.~M. and {Miguel}, Yamila and {Molaverdikhani}, Karan and {Moses}, Julianne I. and {Murray}, Catriona Anne and {Nehring}, Molly and {Nikolov}, Nikolay K. and {Petit dit de la Roche}, D.~J.~M. and {Radica}, Michael and {Roy}, Pierre-Alexis and {Stassun}, Keivan G. and {Taylor}, Jake and {Waalkes}, William C. and {Wachiraphan}, Patcharapol and {Welbanks}, Luis and {Wheatley}, Peter J. and {Aggarwal}, Keshav and {Alam}, Munazza K. and {Banerjee}, Agnibha and {Barstow}, Joanna K. and {Blecic}, Jasmina and {Casewell}, S.~L. and {Changeat}, Quentin and {Chubb}, K.~L. and {Col{\'o}n}, Knicole D. and {Coulombe}, Louis-Philippe and {Daylan}, Tansu and {de Val-Borro}, Miguel and {Decin}, Leen and {Dos Santos}, Leonardo A. and {Flagg}, Laura and {France}, Kevin and {Fu}, Guangwei and {Garc{\'\i}a Mu{\~n}oz}, A. and {Gizis}, John E. and {Glidden}, Ana and {Grant}, David and {Heng}, Kevin and {Henning}, Thomas and {Hong}, Yu-Cian and {Inglis}, Julie and {Iro}, Nicolas and {Kataria}, Tiffany and {Komacek}, Thaddeus D. and {Krick}, Jessica E. and {Lee}, Elspeth K.~H. and {Lewis}, Nikole K. and {Lillo-Box}, Jorge and {Lustig-Yaeger}, Jacob and {Mancini}, Luigi and {Mandell}, Avi M. and {Mansfield}, Megan and {Marley}, Mark S. and {Mikal-Evans}, Thomas and {Morello}, Giuseppe and {Nixon}, Matthew C. and {Ortiz Ceballos}, Kevin and {Piette}, Anjali A.~A. and {Powell}, Diana and {Rackham}, Benjamin V. and {Ramos-Rosado}, Lakeisha and {Rauscher}, Emily and {Redfield}, Seth and {Rogers}, Laura K. and {Roman}, Michael T. and {Roudier}, Gael M. and {Scarsdale}, Nicholas and {Shkolnik}, Evgenya L. and {Southworth}, John and {Spake}, Jessica J. and {Steinrueck}, Maria E. and {Tan}, Xianyu and {Teske}, Johanna K. and {Tremblin}, Pascal and {Tsai}, Shang-Min and {Tucker}, Gregory S. and {Turner}, Jake D. and {Valenti}, Jeff A. and {Venot}, Olivia and {Waldmann}, Ingo P. and {Wallack}, Nicole L. and {Zhang}, Xi and {Zieba}, Sebastian},
        title = "{Identification of carbon dioxide in an exoplanet atmosphere}",
      journal = {\nat},
         year = 2023,
        month = feb,
       volume = {614},
       number = {7949},
        pages = {649-652},
          doi = {10.1038/s41586-022-05269-w},
archivePrefix = {arXiv},
       eprint = {2208.11692},
 primaryClass = {astro-ph.EP},
       adsurl = {https://ui.adsabs.harvard.edu/abs/2023Natur.614..649J}
}

@ARTICLE{Rustamkulov-2023,
       author = {{Rustamkulov}, Z. and {Sing}, D.~K. and {Mukherjee}, S. and {May}, E.~M. and {Kirk}, J. and {Schlawin}, E. and {Line}, M.~R. and {Piaulet}, C. and {Carter}, A.~L. and {Batalha}, N.~E. and {Goyal}, J.~M. and {L{\'o}pez-Morales}, M. and {Lothringer}, J.~D. and {MacDonald}, R.~J. and {Moran}, S.~E. and {Stevenson}, K.~B. and {Wakeford}, H.~R. and {Espinoza}, N. and {Bean}, J.~L. and {Batalha}, N.~M. and {Benneke}, B. and {Berta-Thompson}, Z.~K. and {Crossfield}, I.~J.~M. and {Gao}, P. and {Kreidberg}, L. and {Powell}, D.~K. and {Cubillos}, P.~E. and {Gibson}, N.~P. and {Leconte}, J. and {Molaverdikhani}, K. and {Nikolov}, N.~K. and {Parmentier}, V. and {Roy}, P. and {Taylor}, J. and {Turner}, J.~D. and {Wheatley}, P.~J. and {Aggarwal}, K. and {Ahrer}, E. and {Alam}, M.~K. and {Alderson}, L. and {Allen}, N.~H. and {Banerjee}, A. and {Barat}, S. and {Barrado}, D. and {Barstow}, J.~K. and {Bell}, T.~J. and {Blecic}, J. and {Brande}, J. and {Casewell}, S. and {Changeat}, Q. and {Chubb}, K.~L. and {Crouzet}, N. and {Daylan}, T. and {Decin}, L. and {D{\'e}sert}, J. and {Mikal-Evans}, T. and {Feinstein}, A.~D. and {Flagg}, L. and {Fortney}, J.~J. and {Harrington}, J. and {Heng}, K. and {Hong}, Y. and {Hu}, R. and {Iro}, N. and {Kataria}, T. and {Kempton}, E.~M. -R. and {Krick}, J. and {Lendl}, M. and {Lillo-Box}, J. and {Louca}, A. and {Lustig-Yaeger}, J. and {Mancini}, L. and {Mansfield}, M. and {Mayne}, N.~J. and {Miguel}, Y. and {Morello}, G. and {Ohno}, K. and {Palle}, E. and {Petit dit de la Roche}, D.~J.~M. and {Rackham}, B.~V. and {Radica}, M. and {Ramos-Rosado}, L. and {Redfield}, S. and {Rogers}, L.~K. and {Shkolnik}, E.~L. and {Southworth}, J. and {Teske}, J. and {Tremblin}, P. and {Tucker}, G.~S. and {Venot}, O. and {Waalkes}, W.~C. and {Welbanks}, L. and {Zhang}, X. and {Zieba}, S.},
        title = "{Early Release Science of the exoplanet WASP-39b with JWST NIRSpec PRISM}",
      journal = {\nat},
         year = 2023,
        month = feb,
       volume = {614},
       number = {7949},
        pages = {659-663},
          doi = {10.1038/s41586-022-05677-y},
archivePrefix = {arXiv},
       eprint = {2211.10487},
 primaryClass = {astro-ph.EP},
       adsurl = {https://ui.adsabs.harvard.edu/abs/2023Natur.614..659R}
}

@article{Powell_2024,
  title={Sulphur dioxide in the mid-infrared transmission spectrum of WASP-39b},
  author={Powell, Diana and Feinstein, Adina D and Lee, Elspeth KH and Zhang, Michael and Tsai, Shang-Min and Taylor, Jake and Kirk, James and Bell, Taylor and Barstow, Joanna K and Gao, Peter and others},
  journal={Nature},
volume = {626},
  pages={979-–983},
  year={2024},
  publisher={Nature Publishing Group UK London},
    doi = {10.1038/s41586-024-07040-9}
}

@article{Dong-2023,
author = {Dong, Xiaorui and Pio, Gianmaria and Arafin, Farhan and Laich, Andrew and Baker, Jessica and Ninnemann, Erik and Vasu, Subith S. and Green, William H.},
title = {Butyl Acetate Pyrolysis and Combustion Chemistry: Mechanism Generation and Shock Tube Experiments},
journal = {The Journal of Physical Chemistry A},
volume = {127},
number = {14},
pages = {3231-3245},
year = {2023},
doi = {10.1021/acs.jpca.2c07545}
}

@article{Loyd_2016,
doi = {10.3847/0004-637X/824/2/102},
url = {https://dx.doi.org/10.3847/0004-637X/824/2/102},
year = {2016},
month = {jun},
publisher = {The American Astronomical Society},
volume = {824},
number = {2},
pages = {102},
author = {R. O. P. Loyd and Kevin France and Allison Youngblood and Christian Schneider and Alexander Brown and Renyu Hu and Jeffrey Linsky and Cynthia S. Froning and Seth Redfield and Sarah Rugheimer and Feng Tian},
title = {THE MUSCLES TREASURY SURVEY. III. X-RAY TO INFRARED SPECTRA OF 11 M AND K STARS HOSTING PLANETS},
journal = {\apj}
}

@article{triaud-2015,
  title={WASP-80b has a dayside within the T-dwarf range},
  author={Triaud, Amaury HMJ and Gillon, Micha{\"e}l and Ehrenreich, David and Herrero, Enrique and Lendl, Monika and Anderson, David R and Collier Cameron, Andrew and Delrez, Laetitia and Demory, Brice-Olivier and Hellier, Coel and others},
  journal={Monthly Notices of the Royal Astronomical Society},
  volume={450},
  number={3},
  pages={2279--2290},
  year={2015},
  publisher={The Royal Astronomical Society}
}

@article{RMG-database,
author = {Johnson, Matthew S. and Dong, Xiaorui and Grinberg Dana, Alon and Chung, Yunsie and Farina, David Jr. and Gillis, Ryan J. and Liu, Mengjie and Yee, Nathan W. and Blondal, Katrin and Mazeau, Emily and Grambow, Colin A. and Payne, A. Mark and Spiekermann, Kevin A. and Pang, Hao-Wei and Goldsmith, C. Franklin and West, Richard H. and Green, William H.},
title = {RMG Database for Chemical Property Prediction},
journal = {Journal of Chemical Information and Modeling},
volume = {62},
number = {20},
pages = {4906-4915},
year = {2022},
doi = {10.1021/acs.jcim.2c00965},
}

@Article{Alderson-2023,
  author  = {Alderson, Lili and Wakeford, Hannah R. and Alam, Munazza K. and Batalha, Natasha E. and Lothringer, Joshua D. and Redai, Jea Adams and Barat, Saugata},
  title={Early Release Science of the Exoplanet WASP-39b with JWST NIRSpec G395H},
  journal={Nature},
  year={2023},
  month={Jan},
  day={09},
  issn={1476-4687},
  doi_mute={10.1038/s41586-022-05591-3},
  url_mute={https://doi.org/10.1038/s41586-022-05591-3}
  }

@ARTICLE{Ahrer-2023,
       author = {{Ahrer}, Eva-Maria and {Stevenson}, Kevin B. and {Mansfield}, Megan and {Moran}, Sarah E. and {Brande}, Jonathan and {Morello}, Giuseppe and {Murray}, Catriona A. and {Nikolov}, Nikolay K. and {Petit dit de la Roche}, Dominique J.~M. and {Schlawin}, Everett and {Wheatley}, Peter J. and {Zieba}, Sebastian and {Batalha}, Natasha E. and {Damiano}, Mario and {Goyal}, Jayesh M. and {Lendl}, Monika and {Lothringer}, Joshua D. and {Mukherjee}, Sagnick and {Ohno}, Kazumasa and {Batalha}, Natalie M. and {Battley}, Matthew P. and {Bean}, Jacob L. and {Beatty}, Thomas G. and {Benneke}, Bj{\"o}rn and {Berta-Thompson}, Zachory K. and {Carter}, Aarynn L. and {Cubillos}, Patricio E. and {Daylan}, Tansu and {Espinoza}, N{\'e}stor and {Gao}, Peter and {Gibson}, Neale P. and {Gill}, Samuel and {Harrington}, Joseph and {Hu}, Renyu and {Kreidberg}, Laura and {Lewis}, Nikole K. and {Line}, Michael R. and {L{\'o}pez-Morales}, Mercedes and {Parmentier}, Vivien and {Powell}, Diana K. and {Sing}, David K. and {Tsai}, Shang-Min and {Wakeford}, Hannah R. and {Welbanks}, Luis and {Alam}, Munazza K. and {Alderson}, Lili and {Allen}, Natalie H. and {Anderson}, David R. and {Barstow}, Joanna K. and {Bayliss}, Daniel and {Bell}, Taylor J. and {Blecic}, Jasmina and {Bryant}, Edward M. and {Burleigh}, Matthew R. and {Carone}, Ludmila and {Casewell}, S.~L. and {Changeat}, Quentin and {Chubb}, Katy L. and {Crossfield}, Ian J.~M. and {Crouzet}, Nicolas and {Decin}, Leen and {D{\'e}sert}, Jean-Michel and {Feinstein}, Adina D. and {Flagg}, Laura and {Fortney}, Jonathan J. and {Gizis}, John E. and {Heng}, Kevin and {Iro}, Nicolas and {Kempton}, Eliza M. -R. and {Kendrew}, Sarah and {Kirk}, James and {Knutson}, Heather A. and {Komacek}, Thaddeus D. and {Lagage}, Pierre-Olivier and {Leconte}, J{\'e}r{\'e}my and {Lustig-Yaeger}, Jacob and {MacDonald}, Ryan J. and {Mancini}, Luigi and {May}, E.~M. and {Mayne}, N.~J. and {Miguel}, Yamila and {Mikal-Evans}, Thomas and {Molaverdikhani}, Karan and {Palle}, Enric and {Piaulet}, Caroline and {Rackham}, Benjamin V. and {Redfield}, Seth and {Rogers}, Laura K. and {Roy}, Pierre-Alexis and {Rustamkulov}, Zafar and {Shkolnik}, Evgenya L. and {Sotzen}, Kristin S. and {Taylor}, Jake and {Tremblin}, P. and {Tucker}, Gregory S. and {Turner}, Jake D. and {de Val-Borro}, Miguel and {Venot}, Olivia and {Zhang}, Xi},
        title = "{Early Release Science of the exoplanet WASP-39b with JWST NIRCam}",
      journal = {\nat},
         year = 2023,
        month = feb,
       volume = {614},
       number = {7949},
        pages = {653-658},
          doi = {10.1038/s41586-022-05590-4},
archivePrefix = {arXiv},
       eprint = {2211.10489},
 primaryClass = {astro-ph.EP},
       adsurl = {https://ui.adsabs.harvard.edu/abs/2023Natur.614..653A}
}

@article{Bell-2023,
  title={Methane throughout the atmosphere of the warm exoplanet WASP-80b},
  author={Bell, Taylor J and Welbanks, Luis and Schlawin, Everett and Line, Michael R and Fortney, Jonathan J and Greene, Thomas P and Ohno, Kazumasa and Parmentier, Vivien and Rauscher, Emily and Beatty, Thomas G and others},
  journal={Nature},
  volume={623},
  number={7988},
  pages={709--712},
  year={2023},
  publisher={Nature Publishing Group UK London}
}

@article{yang2024epacris,
doi = {10.3847/1538-4357/ad35c8},
url = {https://dx.doi.org/10.3847/1538-4357/ad35c8},
year = {2024},
month = {may},
publisher = {The American Astronomical Society},
volume = {966},
number = {2},
pages = {189},
author = {Jeehyun Yang and Renyu Hu},
title = {Automated Chemical Reaction Network Generation and Its Application to Exoplanet Atmospheres},
journal = {The Astrophysical Journal}
}

@misc{Benneke-2024_jwst,
      title={JWST Reveals CH$_4$, CO$_2$, and H$_2$O in a Metal-rich Miscible Atmosphere on a Two-Earth-Radius Exoplanet}, 
      author={Björn Benneke and Pierre-Alexis Roy and Louis-Philippe Coulombe and Michael Radica and Caroline Piaulet and Eva-Maria Ahrer and Raymond Pierrehumbert and Joshua Krissansen-Totton and Hilke E. Schlichting and Renyu Hu and Jeehyun Yang and Duncan Christie and Daniel Thorngren and Edward D. Young and Stefan Pelletier and Heather A. Knutson and Yamila Miguel and Thomas M. Evans-Soma and Caroline Dorn and Anna Gagnebin and Jonathan J. Fortney and Thaddeus Komacek and Ryan MacDonald and Eshan Raul and Ryan Cloutier and Lorena Acuna and David Lafrenière and Charles Cadieux and René Doyon and Luis Welbanks and Romain Allart},
      year={2024},
      eprint={2403.03325},
      archivePrefix={arXiv},
      primaryClass={astro-ph.EP}
}

@article{Fortney_2020,
doi = {10.3847/1538-3881/abc5bd},
url = {https://dx.doi.org/10.3847/1538-3881/abc5bd},
year = {2020},
month = {nov},
publisher = {The American Astronomical Society},
volume = {160},
number = {6},
pages = {288},
author = {Jonathan J. Fortney and Channon Visscher and Mark S. Marley and Callie E. Hood and Michael R. Line and Daniel P. Thorngren and Richard S. Freedman and Roxana Lupu},
title = {Beyond Equilibrium Temperature: How the Atmosphere/Interior Connection Affects the Onset of Methane, Ammonia, and Clouds in Warm Transiting Giant Planets},
journal = {The Astronomical Journal}
}

@article{welbanks_2024,
  title={A high internal heat flux and large core in a warm neptune exoplanet},
  author={Welbanks, Luis and Bell, Taylor J and Beatty, Thomas G and Line, Michael R and Ohno, Kazumasa and Fortney, Jonathan J and Schlawin, Everett and Greene, Thomas P and Rauscher, Emily and McGill, Peter and others},
      journal = {Nature},
         year = "2024",
       volume = {630},
        pages ={836--840},
          doi = {https://doi.org/10.1038/s41586-024-07514-w}
}

@article{bryant2025transiting,
  title={A transiting giant planet in orbit around a 0.2-solar-mass host star},
  author={Bryant, Edward M and Jord{\'a}n, Andr{\'e}s and Hartman, Joel D and Bayliss, Daniel and Sedaghati, Elyar and Barkaoui, Khalid and Chouqar, Jamila and Pozuelos, Francisco J and Thorngren, Daniel P and Timmermans, Mathilde and others},
  journal={Nature Astronomy},
  volume={9},
  number={7},
  pages={1031--1044},
  year={2025},
  publisher={Nature Publishing Group UK London}
}

@article{kempton2018framework,
  title={A framework for prioritizing the TESS planetary candidates most amenable to atmospheric characterization},
  author={Kempton, Eliza M-R and Bean, Jacob L and Louie, Dana R and Deming, Drake and Koll, Daniel DB and Mansfield, Megan and Christiansen, Jessie L and L{\'o}pez-Morales, Mercedes and Swain, Mark R and Zellem, Robert T and others},
  journal={Publications of the Astronomical Society of the Pacific},
  volume={130},
  number={993},
  pages={114401},
  year={2018},
  publisher={The Astronomical Society of the Pacific}
}

@article{morales2019giant,
  title={A giant exoplanet orbiting a very-low-mass star challenges planet formation models},
  author={Morales, Juan Carlos and Mustill, AJ and Ribas, I and Davies, MB and Reiners, A and Bauer, FF and Kossakowski, D and Herrero, Enrique and Rodr{\'\i}guez, E and L{\'o}pez-Gonz{\'a}lez, MJ and others},
  journal={Science},
  volume={365},
  number={6460},
  pages={1441--1445},
  year={2019},
  publisher={American Association for the Advancement of Science}
}

@article{xuan2026compositions,
  title={The compositions of the HR 8799 planets reflect accretion of both solids and metal-enriched gas},
  author={Xuan, Jerry W and Ruffio, Jean-Baptiste and Chachan, Yayaati and Ohno, Kazumasa and Kesseli, Aurora and Murray-Clay, Ruth and Lee, Eve J and Moses, Julianne I and Balmer, William O and Baburaj, Aneesh and others},
  journal={The Astrophysical Journal},
  volume={1000},
  number={1},
  pages={27},
  year={2026},
  publisher={The American Astronomical Society}
}

@article{dyrek2024so2,
  title={SO2, silicate clouds, but no CH4 detected in a warm Neptune},
  author={Dyrek, Achr{\`e}ne and Min, Michiel and Decin, Leen and Bouwman, Jeroen and Crouzet, Nicolas and Molli{\`e}re, Paul and Lagage, Pierre-Olivier and Konings, Thomas and Tremblin, Pascal and G{\"u}del, Manuel and others},
  journal={Nature},
  volume={625},
  number={7993},
  pages={51--54},
  year={2024},
  publisher={Nature Publishing Group UK London}
}

@article{triantafillides2026identification,
  title={The Identification of CS2 and Evidence for Carbon-Sulfur Chemical Coupling in a Warm Giant Exoplanet Atmosphere},
  author={Triantafillides, Anastasia and Beatty, Thomas G and Nixon, Matthew C and Bell, Taylor J and Schlawin, Everett and Welbanks, Luis and Greene, Thomas P and Soares-Furtado, Melinda and Line, Jonathan J and Mehta, Nishil and others},
  journal={arXiv preprint arXiv:2604.13168},
  year={2026}
}

@article{yang2024chemical,
  title={Chemical Mapping of Temperate Sub-Neptune Atmospheres: Constraining the Deep Interior \ce{H2O}/\ce{H2} Ratio from the Atmospheric \ce{CO2}/\ce{CH4} Ratio},
  author={Yang, Jeehyun and Hu, Renyu},
  journal={The Astrophysical Journal Letters},
  volume={971},
  number={2},
  pages={L48},
  year={2024},
  month={aug},
  publisher={IOP Publishing},
  doi = {10.3847/2041-8213/ad6b25}
}

@misc{VULCAN_SNCHO_2025,
  author       = {Tsai, Shang-Min},
  title        = {SNCHO\_photo\_network\_2025.txt},
  year         = {2025},
  howpublished = {\url{https://github.com/shami-EEG/VULCAN/blob/master/thermo/SNCHO_photo_network_2025.txt}},
  note         = {Photochemical reaction network distributed with the VULCAN atmospheric chemistry code; accessed May 2026}
}

@article{Zhang_2019_platon,
doi = {10.1088/1538-3873/aaf5ad},
url = {https://doi.org/10.1088/1538-3873/aaf5ad},
year = {2019},
month = {jan},
publisher = {The Astronomical Society of the Pacific},
volume = {131},
number = {997},
pages = {034501},
author = {Zhang, Michael and Chachan, Yayaati and Kempton, Eliza M.-R. and Knutson, Heather A.},
title = {Forward Modeling and Retrievals with PLATON, a Fast Open-source Tool},
journal = {Publications of the Astronomical Society of the Pacific},
}

@article{wogan2025photochem,
  title={The Open-source Photochem Code: A General Chemical and Climate Model for Interpreting (Exo) Planet Observations},
  author={Wogan, Nicholas F and Batalha, Natasha E and Zahnle, Kevin and Krissansen-Totton, Joshua and Catling, David C and Wolf, Eric T and Robinson, Tyler D and Meadows, Victoria and Arney, Giada and Domagal-Goldman, Shawn},
  journal={The Planetary Science Journal},
  volume={6},
  number={11},
  pages={256},
  year={2025},
  publisher={The American Astronomical Society}
}

@article{mang2026picaso,
  title={PICASO 4.0: Clouds and Photochemistry in Climate Models of Brown Dwarfs and Exoplanets},
  author={Mang, James and Batalha, Natasha E and Morley, Caroline V and Wogan, Nicholas F and Mukherjee, Sagnick and Visscher, Channon and Marley, Mark S and Fortney, Jonathan J and Chubb, Katy L and Gao, Peter and others},
  journal={The Astrophysical Journal},
  volume={1000},
  number={1},
  pages={98},
  year={2026},
  publisher={The American Astronomical Society}
}

@article{Bello-Arufe_2026,
doi = {10.3847/1538-3881/ae4fba},
url = {https://doi.org/10.3847/1538-3881/ae4fba},
year = {2026},
month = {may},
publisher = {The American Astronomical Society},
volume = {171},
number = {6},
pages = {354},
author = {Bello-Arufe, Aaron and Hu, Renyu and Zilinskas, Mantas and Yang, Jeehyun and Tokadjian, Armen and Welbanks, Luis and Fu, Guangwei and Greklek-McKeon, Michael and Damiano, Mario and Gomez Barrientos, Jonathan and Knutson, Heather A. and Sing, David K. and Zhang, Xi},
title = {Methane on the Temperate Exo-Saturn TOI-199 b},
journal = {The Astronomical Journal}
}

@article{livingston2026young,
  title={A young progenitor for the most common planetary systems in the Galaxy},
  author={Livingston, John H and Petigura, Erik A and David, Trevor J and Masuda, Kento and Owen, James and Nesvorn{\`y}, David and Batygin, Konstantin and de Leon, Jerome and Mori, Mayuko and Ikuta, Kai and others},
  journal={Nature},
  volume={649},
  number={8096},
  pages={310--314},
  year={2026},
}

@article{mancini2018gaps,
  title={The GAPS programme with HARPS-N at TNG-XVI. Measurement of the Rossiter--McLaughlin effect of transiting planetary systems HAT-P-3, HAT-P-12, HAT-P-22, WASP-39, and WASP-60},
  author={Mancini, L and Esposito, M and Covino, E and Southworth, J and Biazzo, K and Bruni, I and Ciceri, Simona and Evans, D and Lanza, AF and Poretti, E and others},
  journal={Astronomy \& Astrophysics},
  volume={613},
  pages={A41},
  year={2018},
  publisher={Edp Sciences}
}

@article{David_2019_v1298tau_system,
doi = {10.3847/2041-8213/ab4c99},
url = {https://doi.org/10.3847/2041-8213/ab4c99},
year = {2019},
month = {oct},
publisher = {The American Astronomical Society},
volume = {885},
number = {1},
pages = {L12},
author = {David, Trevor J. and Petigura, Erik A. and Luger, Rodrigo and Foreman-Mackey, Daniel and Livingston, John H. and Mamajek, Eric E. and Hillenbrand, Lynne A.},
title = {Four Newborn Planets Transiting the Young Solar Analog V1298 Tau},
journal = {The Astrophysical Journal Letters},
}

@article{suarez2022rapid,
  title={Rapid contraction of giant planets orbiting the 20-million-year-old star V1298 Tau},
  author={Su{\'a}rez Mascare{\~n}o, A and Damasso, Mario and Lodieu, Nicolas and Sozzetti, Alessandro and B{\'e}jar, Victor JS and Benatti, SERENA and Zapatero Osorio, Mar{\'\i}a Rosa and Micela, Giusi and Rebolo, R and Desidera, S and others},
  journal={Nature Astronomy},
  volume={6},
  number={2},
  pages={232--240},
  year={2022},
  publisher={Nature Publishing Group UK London}
}

@ARTICLE{mukherjee_picaso2023,
       author = {{Mukherjee}, Sagnick and {Batalha}, Natasha E. and {Fortney}, Jonathan J. and {Marley}, Mark S.},
        title = "{PICASO 3.0: A One-dimensional Climate Model for Giant Planets and Brown Dwarfs}",
      journal = {\apj},
         year = 2023,
        month = jan,
       volume = {942},
       number = {2},
          eid = {71},
        pages = {71},
          doi = {10.3847/1538-4357/ac9f48},
archivePrefix = {arXiv},
       eprint = {2208.07836},
 primaryClass = {astro-ph.EP},
       adsurl = {https://ui.adsabs.harvard.edu/abs/2023ApJ...942...71M}
}

@article{moses2014chemical,
  title={Chemical kinetics on extrasolar planets},
  author={Moses, Julianne I},
  journal={Philosophical Transactions of the Royal Society A: Mathematical, Physical and Engineering Sciences},
  volume={372},
  number={2014},
  year={2014},
  publisher={The Royal Society}
}

@article{li2012emitted,
  title={Emitted power of Jupiter based on Cassini CIRS and VIMS observations},
  author={Li, Liming and Baines, Kevin H and Smith, Mark A and West, Robert A and P{\'e}rez-Hoyos, Santiago and Trammell, Harold J and Simon-Miller, Amy A and Conrath, Barney J and Gierasch, Peter J and Orton, Glenn S and others},
  journal={Journal of Geophysical Research: Planets},
  volume={117},
  number={E11},
  year={2012},
  publisher={Wiley Online Library}
}

@article{BJORAKER1986579,
title = {The gas composition of jupiter derived from 5-μm airborne spectroscopic observations},
journal = {Icarus},
volume = {66},
number = {3},
pages = {579-609},
year = {1986},
issn = {0019-1035},
doi = {https://doi.org/10.1016/0019-1035(86)90093-X},
url = {https://www.sciencedirect.com/science/article/pii/001910358690093X},
author = {Gordon L. Bjoraker and Harold P. Larson and Virgil G. Kunde},
}

@book{holland2020chemical,
  title={The chemical evolution of the atmosphere and oceans},
  author={Holland, Heinrich D},
  year={2020},
  publisher={Princeton University Press}
}

@book{catling2017atmospheric,
  title={Atmospheric evolution on inhabited and lifeless worlds},
  author={Catling, David C and Kasting, James F},
  year={2017},
  publisher={Cambridge University Press}
}

@inproceedings{moses2024sulfur,
  title={Sulfur photochemistry on warm sub-Neptune and Neptune-class exoplanets},
  author={Moses, Julianne and Tsai, Shang-Min and Fortney, Jonathan and Constantinou, Savvas and Madhusudhan, Nikku and Visscher, Channon and Yu, Xinting and Plane, John and Yang, Jeehyun and Zahnle, Kevin and others},
  booktitle={56th Annual Meeting of the Division for Planetary Sciences},
  volume={56},
  pages={308--06},
  year={2024}
}

@article{green2007predictive,
  title={Predictive kinetics: a new approach for the 21st century},
  author={Green Jr, William H},
  journal={Advances in Chemical Engineering},
  volume={32},
  pages={1--313},
  year={2007},
  publisher={Elsevier}
}

@article{kohse2021combustion,
  title={Combustion in the future: The importance of chemistry},
  author={Kohse-H{\"o}inghaus, Katharina},
  journal={Proceedings of the Combustion Institute},
  volume={38},
  number={1},
  pages={1--56},
  year={2021},
  publisher={Elsevier}
}

@article{miller2021combustion,
  title={Combustion chemistry in the twenty-first century: Developing theory-informed chemical kinetics models},
  author={Miller, James A and Sivaramakrishnan, Raghu and Tao, Yujie and Goldsmith, C Franklin and Burke, Michael P and Jasper, Ahren W and Hansen, Nils and Labbe, Nicole J and Glarborg, Peter and Z{\'a}dor, Judit},
  journal={Progress in Energy and Combustion Science},
  volume={83},
  pages={100886},
  year={2021},
  publisher={Elsevier}
}

@article{green1996deperturbation,
  title={A deperturbation analysis of the B 3$\Sigma$ u-(v′= 0--6) and th B ″3$\Pi$ u (v′= 2--12) states of S2},
  author={Green, ME and Western, CM},
  journal={The Journal of chemical physics},
  volume={104},
  number={3},
  pages={848--864},
  year={1996},
  publisher={American Institute of Physics}
}

@article{green1997upper,
  title={Upper vibrational states of the B ″3 $\Pi$ u state of 32 S 2},
  author={Green, ME and Western, CM},
  journal={Journal of the Chemical Society, Faraday Transactions},
  volume={93},
  number={3},
  pages={365--372},
  year={1997},
  publisher={Royal Society of Chemistry}
}

@article{hrodmarsson2023photodissociation,
  title={Photodissociation and photoionization of molecules of astronomical interest-Updates to the Leiden photodissociation and photoionization cross section database},
  author={Hrodmarsson, HR and Van Dishoeck, EF},
  journal={Astronomy \& Astrophysics},
  volume={675},
  pages={A25},
  year={2023},
  publisher={EDP Sciences}
}

@article{stark2018fourier,
  title={Fourier-transform-spectroscopic photoabsorption cross sections and oscillator strengths for the S2 B$\Sigma$u- 3- X$\Sigma$g- 3 system},
  author={Stark, Glenn and Herde, H and Lyons, JR and Heays, AN and De Oliveira, N and Nave, G and Lewis, BR and Gibson, ST},
  journal={The Journal of chemical physics},
  volume={148},
  number={24},
  year={2018},
  publisher={AIP Publishing}
}

@phdthesis{hull2020collisional,
  title={Collisional transfer between excited electronic states as a mechanism for sulfur mass-independent fractionation},
  author={Hull, Alexander William},
  year={2020},
  school={Massachusetts Institute of Technology}
}

@article{plyler1947infrared,
  title={Infrared absorption spectrum of carbon disulfide},
  author={Plyler, Earle K and Humphreys, CJ},
  journal={J. Res. Natl. Bur. Stand.},
  volume={39},
  number={1},
  pages={59},
  year={1947}
}

@article{barat2025metal,
  title={A metal-poor atmosphere with a hot interior for a young sub-Neptune progenitor: JWST/NIRSpec transmission spectrum of V1298 Tau b},
  author={Barat, Saugata and D{\'e}sert, Jean-Michel and Mukherjee, Sagnick and Goyal, Jayesh M and Xue, Qiao and Kawashima, Yui and Vazan, Allona and Misener, William and Schlichting, Hilke E and Fortney, Jonathan J and others},
  journal={The Astronomical Journal},
  volume={170},
  number={3},
  pages={165},
  year={2025},
  publisher={The American Astronomical Society}
}

@article{allard-2012-phoenix,
    author = {Allard, F. and Homeier, D. and Freytag, B.},
    title = {Models of very-low-mass stars, brown dwarfs and exoplanets},
    journal = {Philosophical Transactions of the Royal Society A: Mathematical, Physical and Engineering Sciences},
    volume = {370},
    number = {1968},
    pages = {2765-2777},
    year = {2012},
    month = {06},
    doi = {10.1098/rsta.2011.0269},
}

@article{Zhang_2025_platon,
doi = {10.3847/1538-3881/ad8cd2},
url = {https://doi.org/10.3847/1538-3881/ad8cd2},
year = {2024},
month = {dec},
publisher = {The American Astronomical Society},
volume = {169},
number = {1},
pages = {38},
author = {Zhang, Michael and Paragas, Kimberly and Bean, Jacob L. and Yeung, Joseph and Chachan, Yayaati and Greene, Thomas P. and Lunine, Jonathan and Deming, Drake},
title = {Retrievals on NIRCam Transmission and Emission Spectra of HD 189733b with PLATON 6, a GPU Code for the JWST Era},
journal = {The Astronomical Journal},
}

@article{GORDON_2026_Hitran,
title = {The HITRAN2024 molecular spectroscopic database},
journal = {Journal of Quantitative Spectroscopy and Radiative Transfer},
volume = {353},
pages = {109807},
year = {2026},
issn = {0022-4073},
doi = {https://doi.org/10.1016/j.jqsrt.2026.109807},
url = {https://www.sciencedirect.com/science/article/pii/S0022407326000014},
author = {I.E. Gordon and L.S. Rothman and R.J. Hargreaves and F.M. Gomez and T. Bertin and C. Hill and R.V. Kochanov and Y. Tan and P. Wcisło and V. Yu. Makhnev and P.F. Bernath and M. Birk and V. Boudon and A. Campargue and A. Coustenis and B.J. Drouin and R.R. Gamache and J.T. Hodges and D. Jacquemart and E.J. Mlawer and A.V. Nikitin and V.I. Perevalov and M. Rotger and S. Robert and J. Tennyson and G.C. Toon and H. Tran and V.G. Tyuterev and E.M. Adkins and A. Barbe and D.M. Bailey and K. Bielska and L. Bizzocchi and T.A. Blake and C.A. Bowesman and P. Cacciani and P. Čermák and A.G. Császár and L. Denis and S.C. Egbert and O. Egorov and A. Yu. Ermilov and A.J. Fleisher and H. Fleurbaey and A. Foltynowicz and T. Furtenbacher and M. Germann and E.R. Guest and J.J. Harrison and J.-M. Hartmann and A. Hjältén and S.-M. Hu and X. Huang and T.J. Johnson and H. Jóźwiak and S. Kassi and M.V. Khan and F. Kwabia-Tchana and T.J. Lee and D. Lisak and A.-W. Liu and O.M. Lyulin and N.A. Malarich and L. Manceron and A.A. Marinina and S.T. Massie and J. Mascio and E.S. Medvedev and V.V. Meshkov and G. Ch. Mellau and M. Melosso and S.N. Mikhailenko and D. Mondelain and H.S.P. Müller and M. O’Donnell and A. Owens and A. Perrin and O.L. Polyansky and P.L. Raston and Z.D. Reed and M. Rey and C. Richard and G.B. Rieker and C. Röske and S.W. Sharpe and E. Starikova and N. Stolarczyk and A.V. Stolyarov and K. Sung and F. Tamassia and J. Terragni and V.G. Ushakov and S. Vasilchenko and B. Vispoel and K.L. Vodopyanov and G. Wagner and S. Wójtewicz and S.N. Yurchenko and N.F. Zobov},
}

@article{yang2026jupiter,
doi = {10.3847/PSJ/ae28d5},
url = {https://doi.org/10.3847/PSJ/ae28d5},
year = {2026},
month = {jan},
publisher = {The American Astronomical Society},
volume = {7},
number = {1},
pages = {2},
author = {Yang, Jeehyun and Hyder, Ali and Hu, Renyu and Lunine, Jonathan I.},
title = {Coupled 1D Chemical Kinetic Transport and 2D Hydrodynamic Modeling Supports a Modest 1–1.5× Supersolar Oxygen Abundance in Jupiter’s Atmosphere},
journal = {The Planetary Science Journal}
}

@article{Crossfield_2025_SO2_shorline,
doi = {10.3847/1538-4357/ae17cb},
url = {https://doi.org/10.3847/1538-4357/ae17cb},
year = {2025},
month = {nov},
publisher = {The American Astronomical Society},
volume = {994},
number = {2},
pages = {184},
author = {Crossfield, Ian J. M. and Ahrer, Eva-Maria and Brande, Jonathan and Kreidberg, Laura and Lothringer, Joshua and Piaulet-Ghorayeb, Caroline and Polman, Jesse and Welbanks, Luis and Kirk, James and Powell, Diana and Khorshid, Niloofar},
title = {Mapping the SO2 Shoreline in Gas Giant Exoplanets},
journal = {The Astrophysical Journal},
}

@article{mukherjee2025effects,
  title={Effects of Planetary Parameters on Disequilibrium Chemistry in Irradiated Planetary Atmospheres: From Gas Giants to Sub-Neptunes},
  author={Mukherjee, Sagnick and Fortney, Jonathan J and Wogan, Nicholas F and Sing, David K and Ohno, Kazumasa},
  journal={The Astrophysical Journal},
  volume={985},
  number={2},
  pages={209},
  year={2025},
  publisher={The American Astronomical Society}
}

@misc{dai2026photochemicalcs2gasdetected,
      title={Photochemical CS$_2$ Gas Detected on a 20-Myr-old Exoplanet}, 
      author={Fei Dai and Erik Petigura and John Livingston and Nicholas Wogan and Sagnick Mukherjee and Zhecheng Hu and Ian J. M. Crossfield and James Owen and Kento Masuda},
      year={2026},
      eprint={2606.00974},
      archivePrefix={arXiv},
      primaryClass={astro-ph.EP},
      url={https://arxiv.org/abs/2606.00974}, 
}

@article{Crossfield_2023,
doi = {10.3847/2041-8213/ace35f},
url = {https://doi.org/10.3847/2041-8213/ace35f},
year = {2023},
month = {jul},
publisher = {The American Astronomical Society},
volume = {952},
number = {1},
pages = {L18},
author = {Crossfield, Ian J. M.},
title = {Volatile-to-sulfur Ratios Can Recover a Gas Giant’s Accretion History},
journal = {The Astrophysical Journal Letters},
}

@article{Veillet-2026,
	author = {{Veillet, R.} and {Venot, O.} and {Sirjean, B.} and {Citrangolo Destro, F.} and {Fournet, R.} and {Al-Refaie, A.} and {Hébrard, E.} and {Glaude, P.-A.} and {Bounaceur, R.}},
	title = {Development of a C/H/O/N/S chemical network: Experimental benchmark, application to exoplanets, and identification of key C/S coupling pathways},
	DOI= "10.1051/0004-6361/202555595",
	url= "https://doi.org/10.1051/0004-6361/202555595",
	journal = {A\&A},
	year = 2026,
	volume = 706,
	pages = "A260",
}

@article{david2019warm,
  title={A warm Jupiter-sized planet transiting the pre-main-sequence star V1298 Tau},
  author={David, Trevor J and Cody, Ann Marie and Hedges, Christina L and Mamajek, Eric E and Hillenbrand, Lynne A and Ciardi, David R and Beichman, Charles A and Petigura, Erik A and Fulton, Benjamin J and Isaacson, Howard T and others},
  journal={The Astronomical Journal},
  volume={158},
  number={2},
  pages={79},
  year={2019},
  publisher={The American Astronomical Society}
}

@article{Zahnle_2009,
doi = {10.1088/0004-637X/701/1/L20},
url = {https://doi.org/10.1088/0004-637X/701/1/L20},
year = {2009},
month = {jul},
publisher = {The American Astronomical Society},
volume = {701},
number = {1},
pages = {L20},
author = {Zahnle, K. and Marley, M. S. and Freedman, R. S. and Lodders, K. and Fortney, J. J.},
title = {ATMOSPHERIC SULFUR PHOTOCHEMISTRY ON HOT JUPITERS},
journal = {The Astrophysical Journal},
}

@article{atreya2020deep,
  title={Deep atmosphere composition, structure, origin, and exploration, with particular focus on critical in situ science at the icy giants},
  author={Atreya, Sushil K and Hofstadter, Mark H and In, Joong Hyun and Mousis, Olivier and Reh, Kim and Wong, Michael H},
  journal={Space Science Reviews},
  volume={216},
  number={1},
  pages={18},
  year={2020},
  publisher={Springer}
}

@article{turrini2021tracing,
  title={Tracing the formation history of giant planets in protoplanetary disks with carbon, oxygen, nitrogen, and sulfur},
  author={Turrini, Diego and Schisano, Eugenio and Fonte, Sergio and Molinari, Sergio and Politi, Romolo and Fedele, Davide and Pani{\'c}, O and Kama, Mihkel and Changeat, Quentin and Tinetti, Giovanna},
  journal={The Astrophysical Journal},
  volume={909},
  number={1},
  pages={40},
  year={2021},
  publisher={The American Astronomical Society}
}

@article{pacetti2022chemical,
  title={Chemical diversity in protoplanetary disks and its impact on the formation history of giant planets},
  author={Pacetti, Elenia and Turrini, Diego and Schisano, Eugenio and Molinari, Sergio and Fonte, Sergio and Politi, Romolo and Hennebelle, Patrick and Klessen, Ralf and Testi, Leonardo and Lebreuilly, Ugo},
  journal={The Astrophysical Journal},
  volume={937},
  number={1},
  pages={36},
  year={2022},
  publisher={The American Astronomical Society}
}

@article{STAGNI2022136723,
title = {An experimental, theoretical and kinetic-modeling study of hydrogen sulfide pyrolysis and oxidation},
journal = {Chemical Engineering Journal},
volume = {446},
pages = {136723},
year = {2022},
issn = {1385-8947},
doi = {https://doi.org/10.1016/j.cej.2022.136723},
url = {https://www.sciencedirect.com/science/article/pii/S1385894722022185},
author = {Alessandro Stagni and Suphaporn Arunthanayothin and Luna {Pratali Maffei} and Olivier Herbinet and Frédérique Battin-Leclerc and Tiziano Faravelli}
}

@misc{zhang2026cnos,
      title={C, N, O, S, and photochemistry in a temperate giant planet orbiting a late M dwarf}, 
      author={Michael Zhang and Qiao Xue and Jeehyun Yang and Vighnesh Nagpal and Michael R. Line and Guangwei Fu and Matthew C. Nixon and Jacob L. Bean and Peter Gao and Eliza M. -R. Kempton and Luis Welbanks and Edward M. Bryant and Daniel Bayliss and Madison Brady and Jean-Michel Désert and Vincent Van Eylen and Jonathan J. Fortney and Andrés Jordán and Vivien Parmentier and Caroline Piaulet-Ghorayeb and Elyar Sedaghati and Kevin B. Stevenson and Amaury H. M. J. Triaud},
      year={2026},
      eprint={2606.23774},
      archivePrefix={arXiv},
      primaryClass={astro-ph.EP},
      url={https://arxiv.org/abs/2606.23774}, 
}

@article{Agundez_2006,
doi = {10.1086/506313},
url = {https://doi.org/10.1086/506313},
year = {2006},
month = {oct},
publisher = {},
volume = {650},
number = {1},
pages = {374},
author = {Agúndez, Marcelino and Cernicharo, José},
title = {Oxygen Chemistry in the Circumstellar Envelope of the Carbon-Rich Star IRC +10216},
journal = {The Astrophysical Journal}
}

@incollection{herbst2008chemistry,
  title={The chemistry of cold interstellar cloud cores},
  author={Herbst, Eric and Millar, Tom J},
  booktitle={Low Temperatures and cold molecules},
  pages={1--54},
  year={2008},
  publisher={World Scientific}
}

@article{millar1990organo,
  title={Organo-sulphur chemistry in dense interstellar clouds},
  author={Millar, TJ and Herbst, E},
  journal={Astronomy and Astrophysics (ISSN 0004-6361), vol. 231, no. 2, May 1990, p. 466-472. Research supported by SERC.},
  volume={231},
  pages={466--472},
  year={1990}
}

@article{vanDishoeck2023,
    author = {van Dishoeck, Ewine F. and Grant, S. and Tabone, B. and van Gelder, M. and Francis, L. and Tychoniec, L. and Bettoni, G. and Arabhavi, A. M. and Gasman, D. and Nazari, P. and Vlasblom, M. and Kavanagh, P. and Christiaens, V. and Klaassen, P. and Beuther, H. and Henning, Th. and Kamp, I.},
    title = {The diverse chemistry of protoplanetary disks as revealed by JWST},
    journal = {Faraday Discussions},
    volume = {245},
    pages = {52-79},
    year = {2023},
    month = {09},
    issn = {1359-6640},
    doi = {10.1039/d3fd00010a},
    url = {https://doi.org/10.1039/d3fd00010a},
    eprint = {https://pubs.rsc.org/fd/article-pdf/doi/10.1039/d3fd00010a/9106630/d3fd00010a.pdf},
}

@article{vidal2017reservoir,
  title={On the reservoir of sulphur in dark clouds: chemistry and elemental abundance reconciled},
  author={Vidal, Thomas HG and Loison, Jean-Christophe and Jaziri, Adam Yassin and Ruaud, Maxime and Gratier, Pierre and Wakelam, Valentine},
  journal={Monthly Notices of the Royal Astronomical Society},
  volume={469},
  number={1},
  pages={435--447},
  year={2017},
  publisher={Oxford University Press}
}

@article{laas2019modeling,
  title={Modeling sulfur depletion in interstellar clouds},
  author={Laas, Jacob C and Caselli, Paola},
  journal={Astronomy \& Astrophysics},
  volume={624},
  pages={A108},
  year={2019},
  publisher={EDP Sciences}
}

@article{oberg2021astrochemistry,
  title={Astrochemistry and compositions of planetary systems},
  author={{\"O}berg, Karin I and Bergin, Edwin A},
  journal={Physics Reports},
  volume={893},
  pages={1--48},
  year={2021},
  publisher={Elsevier}
}

@article{fortney2013framework,
  title={A framework for characterizing the atmospheres of low-mass low-density transiting planets},
  author={Fortney, Jonathan J and Mordasini, Christoph and Nettelmann, Nadine and Kempton, Eliza M-R and Greene, Thomas P and Zahnle, Kevin},
  journal={The Astrophysical Journal},
  volume={775},
  number={1},
  pages={80},
  year={2013},
  publisher={The American Astronomical Society}
}

@article{thorngren2016mass,
  title={The mass--metallicity relation for giant planets},
  author={Thorngren, Daniel P and Fortney, Jonathan J and Murray-Clay, Ruth A and Lopez, Eric D},
  journal={The Astrophysical Journal},
  volume={831},
  number={1},
  pages={64},
  year={2016},
  publisher={The American Astronomical Society}
}

@article{lodders2021relative,
  title={Relative atomic solar system abundances, mass fractions, and atomic masses of the elements and their isotopes, composition of the solar photosphere, and compositions of the major chondritic meteorite groups},
  author={Lodders, Katharina},
  journal={Space Science Reviews},
  volume={217},
  number={3},
  pages={44},
  year={2021},
  publisher={Springer}
}

@article{Reed_2024,
doi = {10.3847/2041-8213/ad74da},
url = {https://doi.org/10.3847/2041-8213/ad74da},
year = {2024},
month = {sep},
publisher = {The American Astronomical Society},
volume = {973},
number = {2},
pages = {L38},
author = {Reed, Nathan W. and Shearer, Randall L. and McGlynn, Shawn Erin and Wing, Boswell A. and Tolbert, Margaret A. and Browne, Eleanor C.},
title = {Abiotic Production of Dimethyl Sulfide, Carbonyl Sulfide, and Other Organosulfur Gases via Photochemistry: Implications for Biosignatures and Metabolic Potential},
journal = {The Astrophysical Journal Letters}
}

@misc{baeyens2026phasedependentchemistrywasp43b,
      title={Phase-dependent chemistry of WASP-43 b revealed with a suite of one-, two-, and three-dimensional models}, 
      author={Robin Baeyens and Julianne I. Moses and Jasmina Blecic and Elspeth K. H. Lee and Lucas Teinturier and Shang-Min Tsai and Jeehyun Yang and Jingxuan Yang and Ludmila Carone and Renyu Hu and Sven Kiefer and Anjali A. A. Piette and Taylor J. Bell and Nicolas Crouzet and Ian Dobbs-Dixon and Christiane Helling and Nicolas Iro and Dominic Samra and Olivia Venot and Jean-Michel Désert},
      year={2026},
      eprint={2607.08486},
      archivePrefix={arXiv},
      primaryClass={astro-ph.EP},
      url={https://arxiv.org/abs/2607.08486}, 
}
\bibliographystyle{aasjournalv7}



\end{document}